\documentclass[final]{elsarticle}
\usepackage{graphicx,amssymb}
\journal{Physica A}

\begin{document}
\begin{frontmatter}

\title{Order-from-disorder effect in the exactly solved mixed spin-(1/2, 1) Ising model on fully frustrated triangles-in-triangles lattices\tnoteref{grant}}
\tnotetext[grant]{This work was financially supported by the grant of The Ministry of Education, Science, Research and Sport of the Slovak Republic under the contract No. VEGA~1/0234/12 and by ERDF EU (European Union European regional development fund) grant under the contract ITMS~26220120047 (activity 3.2).}
\author{Jozef Stre\v{c}ka} 
\ead{jozef.strecka@upjs.sk}
\author{Jana \v{C}is\'{a}rov\'{a}}
\ead{jana.kissova@student.upjs.sk}
\address{Department of Theoretical Physics and Astrophysics, Faculty of Science, \\
P. J. \v{S}af\'{a}rik University, Park Angelinum 9, 040 01 Ko\v{s}ice, Slovak Republic}

\begin{abstract}
The mixed spin-(1/2, 1) Ising model on two fully frustrated triangles-in-triangles lattices is exactly solved with the help of the generalized star-triangle transformation, which establishes a rigorous mapping correspondence with the equivalent spin-1/2 Ising model on a triangular lattice. It is shown that the mutual interplay between the spin frustration and single-ion anisotropy gives rise to various spontaneously ordered and disordered ground states, which differ mainly in an occurrence probability of the non-magnetic spin state of the integer-valued decorating spins. We have convincingly evidenced a possible coexistence of the spontaneous long-range order with a partial disorder within the striking ordered-disordered ground state, which manifest itself through a non-trivial criticality at finite temperatures as well. A rather rich critical behaviour including the order-from-disorder effect and reentrant phase transitions with either two or three successive critical points is also found. 
\end{abstract}

\begin{keyword}
Ising model \sep order-from disorder effect \sep reentrance 
\PACS 05.50.+q \sep 75.10.Hk \sep 75.10.Jm 
\end{keyword}

\end{frontmatter}

\section{Introduction}

Two-dimensional frustrated Ising spin systems have been actively studied during the past few decades, because the spin frustration usually manifests itself at low enough temperatures through intriguing phase transitions and magnetic properties (see Refs. \cite{lieb86,diep04} and references therein). The term \textit{spin frustration} is ascribed to an incapability of spins to find an optimal spin configuration, which would simultaneously satisfy all interactions between spins. The spin frustration can arise either from a mutual competition between different interactions or from a geometric topology of the considered lattice. It is worthwhile to remark that several two-dimensional frustrated Ising models have been solved quite rigorously and those exact solutions have brought a deeper understanding into a variety of striking phenomena such as an existence of highly degenerate ground states with a non-zero residual entropy \cite{lieb86} or reentrant phase transitions with a few successive critical points \cite{diep04}. The antiferromagnetic spin-$\frac{1}{2}$ Ising model on a triangular lattice was presumably the first exactly solved model with the macroscopically degenerate ground state having a non-zero residual entropy \cite{hout50,wann50,temp50,husi50,syoz50,newe50,pott52,domb60,step70,cunn74,baxt89}, while the spin-$\frac{1}{2}$ Ising model on a centered square (union jack) lattice with the competing (antiferromagnetic) next-nearest-neighbour interaction was likely the first exactly solved model showing reentrant phase transitions \cite{vaks66,mori86,chik87}. An absence of finite-temperature phase transition above the highly degenerate ground state has been later rigorously proved for the spin-$\frac{1}{2}$ Ising model on a kagom\'e lattice \cite{kano53,kano66}, fully frustrated square, triangular and honeycomb lattices \cite{vill77,wolf82}, brickwork lattice \cite{seze81}, chessboard lattice \cite{wozi82}, pentagon lattice \cite{wald84,wald85}, triangular kagom\'e lattice \cite{zhen05,loh08} and others.  
     
Although it could be naively expected that the macroscopic degeneracy of the ground state ultimately inhibits spontaneous long-range order, some two-dimensional frustrated Ising models with the highly (but not too highly) degenerate ground state exhibit a striking coexistence of spontaneous long-range order with a partial disorder. Indeed, the exact solutions for the spin-$\frac{1}{2}$ Ising model on a centered square lattice \cite{vaks66,mori86,chik87}, piled-up and zig-zag domino lattices \cite{andr79}, centered honeycomb lattice \cite{diep91,diep92} and generalized kagom\'e lattice \cite{diep92,azar87,deba91} have convincingly evidenced a presence of incomplete spontaneous order with a considerable value of the residual (zero-point) entropy. It is noteworthy that the reentrance with a few successive critical points is also observed in most of the aforementioned exactly solved Ising models. Besides, the spin-$\frac{1}{2}$ Ising model on a domino lattice shows a spectacular phenomenon called as \textit{order-from-disorder effect} \cite{vill80,bida81,zitt82}, which refers to an imperfect spontaneous order that is thermally induced just above a disordered ground state. The fact that thermal fluctuations may produce a spontaneous order above a disordered ground state by selecting a state with the largest entropy obviously represents a highly contra-intuitive cooperative phenomenon. 

The mixed-spin Ising models represent probably the most natural extensions of the simple spin-$\frac{1}{2}$ Ising model. The foremost reason for a theoretical investigation of the mixed-spin Ising models closely relates to a more diverse critical behaviour they usually display compared with their single-spin counterparts. Despite a great effort, there are only a few exactly solved examples of the frustrated mixed-spin Ising models on two-dimensional lattices so far. Among these, one could mention the mixed-spin Ising model on a honeycomb lattice \cite{gonc87}, centered square lattice \cite{lipo95,stre06,cano06}, diced lattice \cite{jasc05} and decorated Bethe lattice \cite{stre12}, in which the competing character of the further-neighbour interaction causes an appearance of reentrant phase transitions. The main goal of the present paper is to exactly treat a mixed spin-($\frac{1}{2}$, 1) Ising model on two fully frustrated triangles-in-triangles lattices, which provide a useful playground for investigating phase transitions 
induced by the spin frustration. 

The outline of the present paper is as follows. In Section \ref{model}, the model under investigation will be described in detail and basic steps of the exact solution will be clarified. In Section \ref{result}, the detailed discussion of the most interesting results will be presented. In particular, we will describe the ground-state and finite-temperature phase diagrams, temperature variations of the sublattice magnetizations and quadrupolar moment. The Section \ref{conclusion} brings some conclusions and future outlooks.

\section{Model and method}
\label{model}

\begin{figure}[t]
\begin{center}
\includegraphics[width=11.0cm]{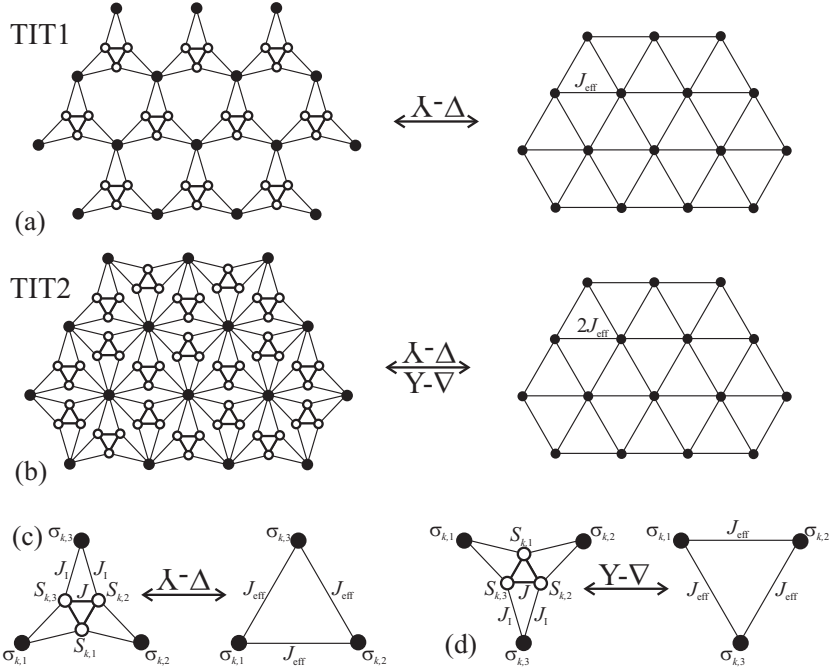}
\vspace{-0.2cm}
\caption{(a)-(b) The mixed spin-($\frac{1}{2}$, 1) Ising model on two considered TIT lattices and its rigorous mapping to the effective spin-$\frac{1}{2}$ Ising model on a triangular lattice. Full circles denote lattice positions of the nodal Ising spins $\sigma=\frac{1}{2}$ and the empty ones lattice positions of the decorating Ising spins $S=1$; (c)-(d) A diagrammatic representation of the star-triangle transformation used for an elementary six-spin star cluster.}
\label{fig1}
\end{center}
\end{figure}

Let us introduce a mixed spin-$\frac{1}{2}$ and spin-$1$ Ising model on two geometrically related TIT lattices, which are schematically depicted in Fig.~\ref{fig1}(a)-(b). As one can  see from this figure, both considered TIT lattices can be derived from a simple triangular lattice by decorating its faces with additional triangles of sites. TIT1 lattice shown in Fig.~\ref{fig1}(a) is obtained by placing an additional triangle of decorating sites into each up-pointing triangle of the underlying triangular lattice, while TIT2 lattice from Fig.~\ref{fig1}(b) is designed by placing an extra triangle of decorating sites into each triangle of the underlying triangular lattice. 
With respect to this, the lattice sites of the underlying triangular lattice will be referred to as \textit{nodal} sites, while the new lattice sites of the additionally placed triangles will be called as \textit{decorating} sites. Suppose furthermore that the nodal and decorating lattice sites are occupied by the Ising spins $\sigma = \frac{1}{2}$ and $S = 1$, respectively. The total Hamiltonian of the mixed spin-($\frac{1}{2}$, 1) Ising model defined on both aforedescribed TIT lattices then reads
\begin{eqnarray}
{\cal H} = - J \sum_{\langle i,j \rangle}^{3 \gamma N} S_i S_j - J_{\rm 1} \sum_{\langle k,l \rangle}^{6 \gamma N} S_k \sigma_l - D \sum_{i=1}^{3 \gamma N} S_i^2,  
\label{Htot} 
\end{eqnarray}
where $\sigma_l = \pm \frac{1}{2}$ and $S_i = \pm 1,0$ are the Ising spin variables placed at the nodal and decorating lattice sites, respectively, $N$ denotes the total number of the nodal lattice sites and $\gamma N$ labels the total number of the decorating triangles (i.e. $\gamma = 1$ for the TIT1 lattice shown in Fig.~\ref{fig1}(a) and $\gamma = 2$ for the TIT2 lattice displayed in Fig.~\ref{fig1}(b)). The parameter $J$ marks the pair interaction between the nearest-neighbour decorating spins, the parameter $J_1$ is the pair interaction between the nearest-neighbour nodal and decorating spins, respectively, and the parameter $D$ stands for the single-ion anisotropy acting on the decorating spins only. For further convenience, it is advisable to rewrite the total Hamiltonian (\ref{Htot}) as a sum over Hamiltonians of the six-spin star clusters schematically illustrated on the left-hand-side of Fig.~\ref{fig1}(c)-(d)
\begin{eqnarray}
{\cal H} = \sum_{k=1}^{\gamma N} {\cal H}_k, 
\label{Hsum}
\end{eqnarray}
with
\begin{eqnarray}
{\cal H}_k = - J \sum_{i=1}^3 S_{k,i} S_{k,i+1} - J_1 \sum_{i=1}^3  S_{k,i} \bigl( \sigma_{k,i} + \sigma_{k,i+1} \bigr) -  D \sum_{i=1}^3  S_{k,i}^2.
\label{Hclu}
\end{eqnarray}
Note that the convention $\sigma_{k,4} \equiv \sigma_{k,1}$ and $S_{k,4} \equiv S_{k,1}$ is used in the definition (\ref{Hclu}) of the cluster Hamiltonian ${\cal H}_k$, which involves all the interaction terms belonging to the $k$th six-spin star cluster. Apparently, the summations over spin degrees of freedom of the decorating spins from different cluster Hamiltonians (\ref{Hclu}) can be performed independently of each other and hence, the partition function of the mixed-spin Ising model on the TIT lattices can be formally written in this compact form  
\begin{eqnarray}
{\cal Z} \!\!\!&=&\!\!\! \sum_{\{ \sigma_i \}} \prod_{k=1}^{\gamma N} \sum_{ S_{k,1} } \sum_{S_{k,2}} \sum_{S_{k,3}} \exp(-\beta {\cal H}_k) 
                = \sum_{\{ \sigma_i \}} \prod_{k=1}^{\gamma N} {\cal Z}_k (\sigma_{k,1}, \sigma_{k,2}, \sigma_{k,3}),
\label{PFF} 
\end{eqnarray}
where $\beta = 1/(k_{\rm B} T)$, $k_{\rm B}$ is Boltzmann's constant, $T$ is the absolute temperature and the relevant product runs over all six-spin star clusters.  After summing up over the spin degrees of freedom of the decorating Ising spins $S_{k,1}$, $S_{k,2}$ and $S_{k,3}$, the effective Boltzmann's weight ${\cal Z}_k$ solely depends on the three nodal Ising spins $\sigma_{k,1}$, $\sigma_{k,2}$ and $\sigma_{k,3}$. Moreover, the explicit form of the relevant Boltzmann's factor automatically suggests a possibility to implement the generalized star-triangle transformation \cite{fish59,roja09,stre10,stre10v}, which is schematically illustrated in Fig.~\ref{fig1}(c)-(d) and mathematically given by 
\begin{eqnarray}
{\cal Z}_k \!\!\!\!\!\!\!\!&&\!\!\!\!\!\!\!\! (\sigma_{k,1}, \sigma_{k,2}, \sigma_{k,3}) = \sum_{ S_{k,1} } \sum_{S_{k,2}} \sum_{S_{k,3}} \exp(-\beta {\cal H}_k) \nonumber \\
\!\!\!&=&\!\!\! 1 + 6 \exp [\beta (-J + 3D)] + 2 \exp [3 \beta (J + D)] \cosh [2 \beta J_1 (\sigma_{k,1} + \sigma_{k,2} + \sigma_{k,3})] \nonumber \\
\!\!\!&+&\!\!\! 2 \exp [\beta (J + 2D)] \bigl \{ \cosh [\beta J_1 (\sigma_{k,1} + \sigma_{k,2} + 2 \sigma_{k,3})] \nonumber \\
\!\!\!&+&\!\!\! \cosh [\beta J_1 (\sigma_{k,1} + 2 \sigma_{k,2} + \sigma_{k,3})] + \cosh [\beta J_1 (2 \sigma_{k,1} + \sigma_{k,2} + \sigma_{k,3})] \bigr \} \nonumber \\
\!\!\!&+&\!\!\! 2 \exp (\beta D) \bigl \{ \cosh [\beta J_1 (\sigma_{k,1} + \sigma_{k,2})] + \cosh [\beta J_1 (\sigma_{k,2} + \sigma_{k,3})] \nonumber \\
\!\!\!&+&\!\!\! \cosh [\beta J_1 (\sigma_{k,1} + \sigma_{k,3})] \bigr \} + 2 \exp [\beta (-J + 2D)] \bigl \{ \cosh [\beta J_1 (\sigma_{k,1} - \sigma_{k,2})] \nonumber \\
\!\!\!&+&\!\!\! \cosh [\beta J_1 (\sigma_{k,2} - \sigma_{k,3})] + \cosh [\beta J_1 (\sigma_{k,1} - \sigma_{k,3})] \bigr \} \nonumber \\
\!\!\!&=&\!\!\! A \exp [\beta J_{\rm eff} (\sigma_{k,1} \sigma_{k,2} + \sigma_{k,2}\sigma_{k,3} + \sigma_{k,3}\sigma_{k,1})].
\label{GAT1} 
\end{eqnarray}
The star-triangle transformation (\ref{GAT1}) actually represents a set of eight equations corresponding to eight different spin configurations of the three nodal Ising spins, but one merely gets just two independent equations from this set (on assumption that the external magnetic field is absent) that unambiguously determine so far unspecified mapping parameters $A$ and $J_{\rm{eff}}$ 
\begin{eqnarray}
A = {\bigl( V_1 V_2^3 \bigr)}^{\frac{1}{4}}, \qquad  \beta J_{\rm{eff}} = \ln \biggl( \frac{V_1}{V_2}\biggr).
\label{TP}
\end{eqnarray}
In above, the expression $V_1$ denotes the effective Boltzmann's weight that corresponds to two particular uniform spin configurations with a parallel orientation of all three nodal Ising spins 
\begin{eqnarray}
V_1 \!\!\!&\equiv&\!\!\! {\cal Z}_k \left(\pm \frac{1}{2}, \pm \frac{1}{2}, \pm\frac{1}{2}\right) = 1 + 6 \exp (\beta D) \cosh (\beta J_1)  \nonumber \\ 
\!\!\!&+&\!\!\! 6 \exp [\beta (J + 2D)] \cosh (2 \beta J_1) +  6 \exp [\beta (-J + 2D)] \nonumber \\
\!\!\!&+&\!\!\! 6 \exp [\beta (-J + 3D)] \cosh (\beta J_1)  + 2 \exp [3 \beta (J + D)] \cosh (3 \beta J_1) ,
\label{V1}
\end{eqnarray}
while the other expression $V_2$ represents the effective Boltzmann's weight corresponding to six non-uniform spin configurations in which one out of three nodal Ising spins is aligned in opposite with respect to the other two
\begin{eqnarray}
V_2 \!\!\!&\equiv&\!\!\!  {\cal Z}_k \left(\pm \frac{1}{2}, \pm \frac{1}{2}, \mp \frac{1}{2}\right) = {\cal Z}_k \left(\pm \frac{1}{2},\mp \frac{1}{2}, \pm \frac{1}{2}\right) = {\cal Z}_k \left(\mp \frac{1}{2},\pm \frac{1}{2}, \pm \frac{1}{2}\right) \nonumber \\
 \!\!\!&=&\!\!\! 1 + 2 \exp (\beta D) [2 + \cosh (\beta J_1)] + 6 \exp [\beta (-J + 3D)] \cosh (\beta J_1) \nonumber \\ 
\!\!\!&+&\!\!\! 4 \exp (2 \beta D) \cosh (\beta J) [1 + 2 \cosh (\beta J_1)] + 2 \exp [3 \beta (J + D)] \cosh (\beta J_1)\!.
\label{V2}
\end{eqnarray}
By inserting the star-triangle transformation (\ref{GAT1}) into the relation (\ref{PFF}) one obtains an exact mapping relationship between the partition function ${\cal Z}$ of the mixed spin-($\frac{1}{2}$, $1$) Ising model on the TIT lattice and respectively, the partition function ${\cal Z}_{\rm IM}$ of the corresponding spin-$\frac{1}{2}$ Ising model on a simple triangular lattice with the effective nearest-neighbour interaction $\gamma J_{\rm{eff}}$
\begin{eqnarray}
{\cal Z} (\beta, J, J_1, D) = A^{\gamma N} {\cal Z}_{\rm{IM}} (\beta, \gamma J_{\rm{eff}}),
\label{MapCor} 
\end{eqnarray}
which is defined through the Hamiltonian
\begin{eqnarray}
{\cal H}_{\rm{IM}} = -\gamma J_{\rm{eff}} \sum_{\langle i,j \rangle}^{3N} \sigma_i \sigma_j.
\label{Hef} 
\end{eqnarray}
Hence, it follows that the partition function of the mixed-spin Ising model on the TIT lattices can be easily obtained from the corresponding exact result for the partition function of the spin-$\frac{1}{2}$ Ising model on a triangular lattice \cite{hout50,wann50,temp50,husi50,syoz50,newe50}. It is quite apparent from Eqs.~(\ref{MapCor}) and (\ref{Hef}) that the only difference in an exact treatment of both investigated TIT lattices lies in a relative strength of the effective coupling of the equivalent spin-$\frac{1}{2}$ Ising model on a triangular lattice, which is two times greater for the TIT2 lattice than that of the TIT1 lattice. Using the established mapping equivalence (\ref{MapCor}) between the partition functions, all basic thermodynamic quantities of the mixed-spin Ising model on the TIT lattices can be readily calculated from the relevant quantities of the effective spin-$\frac{1}{2}$ Ising model on a triangular lattice thoroughly investigated in several previous studies \cite{hout50,wann50,temp50,husi50,syoz50,newe50,pott52,domb60,step70,cunn74,baxt89}.

For illustration, let us calculate spontaneous magnetizations of the nodal and decorating Ising spins, respectively. Adopting the exact mapping theorems developed by Barry \textit{et al}. \cite{barr82,barr88,khat90,barr91,barr95}, the canonical ensemble average of any function of the nodal spins performed within the mixed-spin Ising model on the TIT lattice directly equals to the canonical ensemble average of the same function of the nodal spins in the corresponding spin-$\frac{1}{2}$ Ising model on a triangular lattice
\begin{eqnarray}
\langle f_1 (\sigma_i, \sigma_j, \ldots, \sigma_k) \rangle = \langle f_1 (\sigma_i, \sigma_j, \ldots, \sigma_k) \rangle_{\rm IM}.
\label{bmp}
\end{eqnarray}
It directly follows from Eq.~(\ref{bmp}) that the spontaneous magnetization $m_{\sigma}$ of the nodal spins in the mixed-spin Ising model on the TIT lattice equals to the spontaneous magnetization $m_{\rm{IM}}$ of the equivalent spin-$\frac{1}{2}$ Ising model on a simple triangular lattice with the effective nearest-neighbour interaction $\gamma J_{\rm{eff}}$
\begin{eqnarray}
m_{\sigma} \equiv \langle \sigma_{k,i} \rangle = \langle \sigma_{k,i} \rangle_{\rm IM} \equiv m_{\rm IM} (\gamma J_{\rm eff}).
\label{m0}
\end{eqnarray}
According to this, one may obtain an exact expression for the spontaneous magnetization of the nodal spins from the spontaneous magnetization of the spin-$\frac{1}{2}$ Ising model on a simple triangular lattice \cite{pott52} 
\begin{eqnarray}
m_{\sigma} = m_{\rm IM} = \frac{1}{2} \biggl[ 1 - \frac{16 z^{6}}{(1 + 3z^2)(1 - z^2)^3}\biggr]^{1/8},
\label{mA}
\end{eqnarray}
whereas the parameter $z = \exp(- \gamma \beta J_{\rm{eff}}/2)$ depends just upon the effective coupling unambiguously given by Eq.~(\ref{TP}).

On the other hand, the spontaneous magnetization of the decorating spins can be rigorously calculated by adopting of the generalized Callen-Suzuki identity \cite{call63,suzu65,balc02}, which allows us to get the canonical ensemble average of any function of the decorating spins $S_{k,i}$ ($i=1,2,3$) from the exact spin identity
\begin{eqnarray}
\langle f_2 (S_{k,1}, S_{k,2}, S_{k,3}) \rangle = 
\left \langle \frac{\displaystyle \sum_{S_{k,1}} \displaystyle \sum_{S_{k,2}} \displaystyle \sum_{S_{k,3}} f_2 (S_{k,1}, S_{k,2}, S_{k,3}) \exp (-\beta {\cal H}_k)}
                   {\displaystyle \sum_{S_{k,1}} \displaystyle \sum_{S_{k,2}} \displaystyle \sum_{S_{k,3}} \exp (-\beta {\cal H}_k)} \right \rangle.
\label{CSI}
\end{eqnarray}
After straightforward but a little bit cumbersome calculation based on the exact spin identity (\ref{CSI}), the spontaneous magnetization $m_S$ of the decorating spins in the mixed spin-($\frac{1}{2}$, 1) Ising model on the TIT lattice can be computed from the spontaneous magnetization $m_{\rm {IM}}$ and triplet correlation function $t_{\rm {IM}} \equiv \langle \sigma_{k,1} \sigma_{k,2} \sigma_{k,3} \rangle_{\rm IM}$ of the corresponding spin-$\frac{1}{2}$ Ising model on a triangular lattice 
\begin{eqnarray}
m_S \equiv \langle S_{k,i} \rangle = \frac{m_{\rm IM}}{2} \biggl(3 \frac{W_1}{V_1} + \frac{W_2}{V_2} \biggr)+ 2 t_{\rm {IM}} \biggl( \frac{W_1}{V_1} - \frac{W_2}{V_2} \biggr). 
\label{mB}
\end{eqnarray}
For completeness, let us recall here the relevant exact result for the triplet correlation function $t_{\rm IM} \equiv \langle \sigma_{k,1} \sigma_{k,2} \sigma_{k,3} \rangle_{\rm IM}$ of the spin-$\frac{1}{2}$ Ising model on the triangular lattice \cite{baxt89} 
\begin{eqnarray}
t_{\rm IM} = \frac{m_{\rm IM}}{4} \!\! \left[ 1 + 2 \frac{y - 2 y^{-1} + 1 - \sqrt{(y+3)(y-1)}}{y + y^{-1} - 2} \right]\!\!, \qquad
y = \exp \left(\beta \gamma J_{\rm{eff}} \right)
\label{timco} 
\end{eqnarray}
together with the explicit form of two newly defined parameters $W_1$ and $W_2$  
\begin{eqnarray}
W_1 \!\!\!&=&\!\!\! 2 \exp (\beta D) \sinh (\beta J_1) +  4 \exp [\beta (J + 2D)] \sinh (2 \beta J_1) \nonumber \\ 
\!\!\!&+&\!\!\! 2 \exp [\beta (-J + 3D)] \sinh (\beta J_1) + 2 \exp [3 \beta (J + D)] \sinh (3 \beta J_1), \nonumber \\
W_2 \!\!\!&=&\!\!\! 2 \exp (\beta D) \sinh (\beta J_1) + 8 \exp [\beta (J + 2D)] \sinh (\beta J_1) \nonumber \\ 
\!\!\!&+&\!\!\! 2 \exp [\beta (-J + 3D)] \sinh (\beta J_1) + 6 \exp [3 \beta (J + D)] \sinh (\beta J_1).
\label{W1}
\end{eqnarray}
Next, let us take advantage of the generalized Callen-Suzuki identity (\ref{CSI}) in order to calculate the quadrupolar moment of the decorating spins. Following the procedure worked out previously one is also able to derive the precise formula   
\begin{eqnarray}
q_S \equiv \langle S_{k,i}^2 \rangle = \frac{1}{12} \biggl( \frac{U_1}{V_1} + 3 \frac{U_2}{V_2} \biggr)+ \rho_{\rm {IM}} \biggl( \frac{U_1}{V_1} - \frac{U_2}{V_2} \biggr),
\label{qB}
\end{eqnarray}
according to which the quadrupolar moment is expressed in terms of the nearest-neighbour pair correlation function $\rho_{\rm  {IM}} = \langle \sigma_{k,1} \sigma_{k,2} \rangle_{\rm IM}$ of the corresponding spin-$\frac{1}{2}$ Ising model on a triangular lattice \cite{domb60} and the functions $U_1$, $U_2$ 
\begin{eqnarray}
U_1 \!\!\!&=&\!\!\! 6 \exp(\beta D) \cosh (\beta J_1) + 12 \exp [\beta (J + 2D)] \cosh (2 \beta J_1)   \nonumber \\ 
\!\!\!&+&\!\!\! 12 \exp [\beta (-J + 2D)] + 18 \exp [\beta (-J + 3D)] \cosh (\beta J_1) \nonumber \\
\!\!\!&+&\!\!\! 6 \exp [3 \beta (J + D)] \cosh (3 \beta J_1), \nonumber \\
U_2  \!\!\!&=&\!\!\! 2 \exp (\beta D) [2 + \cosh (\beta J_1)] + 4 \exp [\beta (J + 2D)][1 + 2 \cosh (\beta J_1)] \nonumber \\ 
\!\!\!&+&\!\!\! 4 \exp [\beta (-J + 2D)] [1 + 2 \cosh(\beta J_1)] +   18 \exp [\beta (-J + 3D)] \cosh (\beta J_1) \nonumber \\
\!\!\!&+&\!\!\! 6 \exp [3 \beta (J + D)] \cosh (\beta J_1).
\label{U1U2}
\end{eqnarray}

Last but not least, we will propose a suitable criterion in order to determine critical points of the mixed spin-($\frac{1}{2}$, 1) Ising model on the TIT lattices. It is quite evident from Eqs.~(\ref{mA}), (\ref{mB}) and (\ref{timco}) that the spontaneous magnetizations $m_{\sigma}$ and $m_S$ of the nodal and decorating spins simultaneously vanish just if the spontaneous magnetization $m_{\rm IM}$ of the equivalent spin-$\frac{1}{2}$ Ising model on a triangular lattice becomes zero. This result is taken to mean that the mixed spin-($\frac{1}{2}$, 1) Ising model on the TIT lattices reaches a critical point if and only if the corresponding spin-$\frac{1}{2}$ Ising model on a triangular lattice arrives at a critical point as well. From this point of view, the critical temperature of the mixed spin-($\frac{1}{2}$, 1) Ising model on the TIT lattices can be straightforwardly attained by comparing the effective coupling of the equivalent spin-$\frac{1}{2}$ Ising model on a triangular lattice with its critical value
\begin{eqnarray}
\beta_c \gamma J_{\rm eff} = \ln 3 \qquad \Longleftrightarrow \qquad V_1^{\gamma} (\beta_c) = 3 V_2^{\gamma} (\beta_c),
\label{cc}
\end{eqnarray}
where $\beta_c = 1/(k_{\rm B} T_c)$ and $T_c$ is a critical temperature. The expressions $V_1 (\beta_c)$ and $V_2 (\beta_c)$ denote the Boltzmann's weights given by Eqs.~(\ref{V1}) and (\ref{V2}) except that the inverse critical temperature $\beta_c$ enters the respective formulas instead of $\beta$. 

\section{Results and discussion}
\label{result}

Now, let us proceed to a discussion of the most interesting results for the mixed spin-($\frac{1}{2}$, 1) Ising model on the two considered TIT lattices, which have been derived in the foregoing section. Although all the results derived previously are valid regardless of whether the interaction constants $J$ and $J_1$ are ferromagnetic or antiferromagnetic, henceforth we will restrict our attention only to a detailed analysis of the special case with the ferromagnetic coupling $J_1 > 0$ because the consideration of the antiferromagnetic coupling $J_1 < 0$ would merely cause a trivial change in a relative orientation of the nearest-neighbour nodal and decorating spins. By contrast, the respective change in the coupling $J$ between the nearest-neighbour decorating spins has a significant impact upon the overall magnetic behaviour of the mixed spin-($\frac{1}{2}$, 1) Ising model on the TIT lattices, since the antiferromagnetic interaction $J<0$ is responsible for an emergence of the geometric spin frustration that may demolish a spontaneous long-range order unlike the ferromagnetic interaction $J>0$. With regard to this, we will explore magnetic properties of the mixed spin-($\frac{1}{2}$, 1) Ising model on the TIT lattices by considering both ferromagnetic ($J>0$) as well as antiferromagnetic ($J<0$) interaction between the nearest-neighbouring decorating spins, whereas the ferromagnetic interaction $J_1>0$ will serve as the energy unit. 

\begin{figure}[t]
\includegraphics[width=0.57\textwidth]{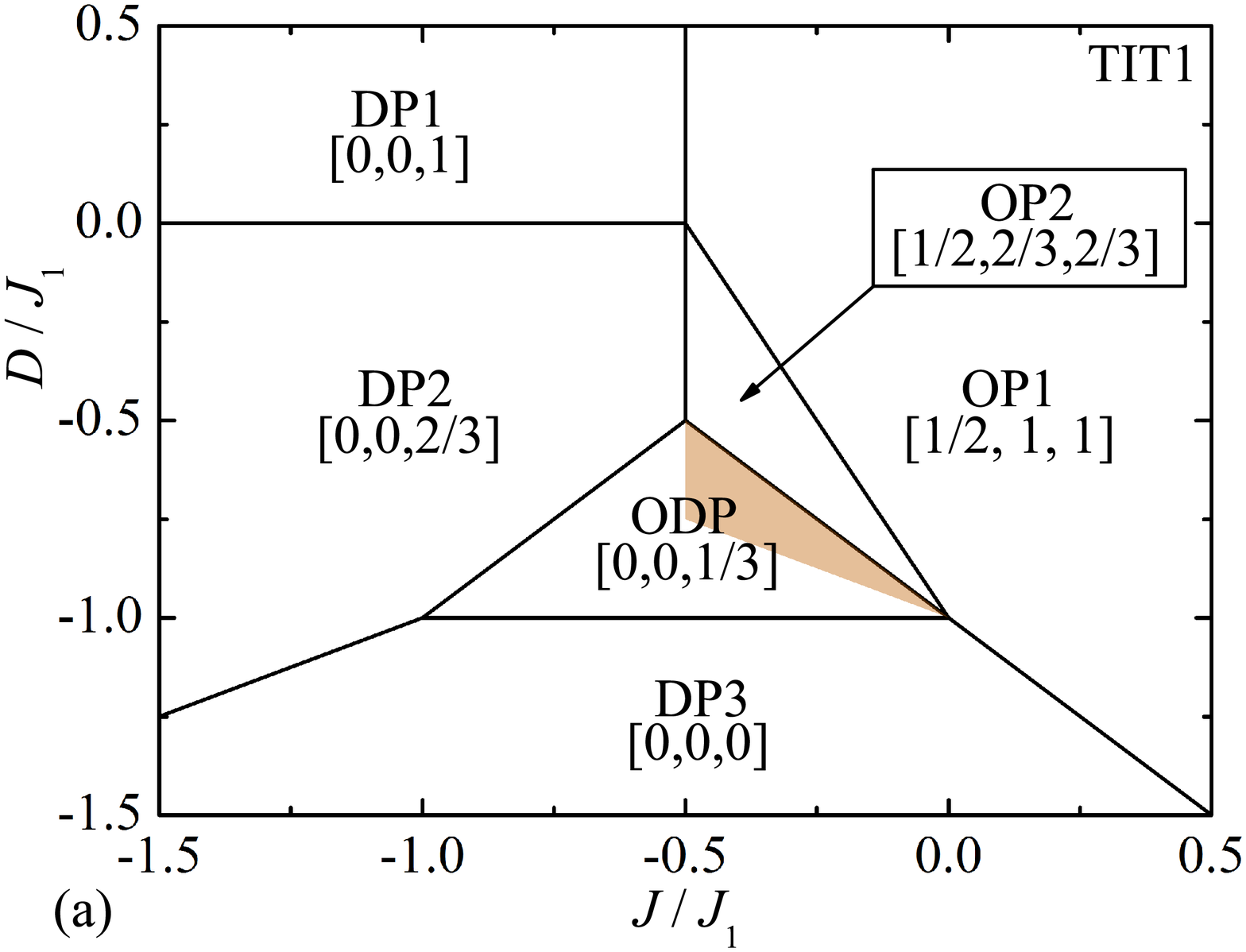}
\hspace{-1.0cm}
\includegraphics[width=0.57\textwidth]{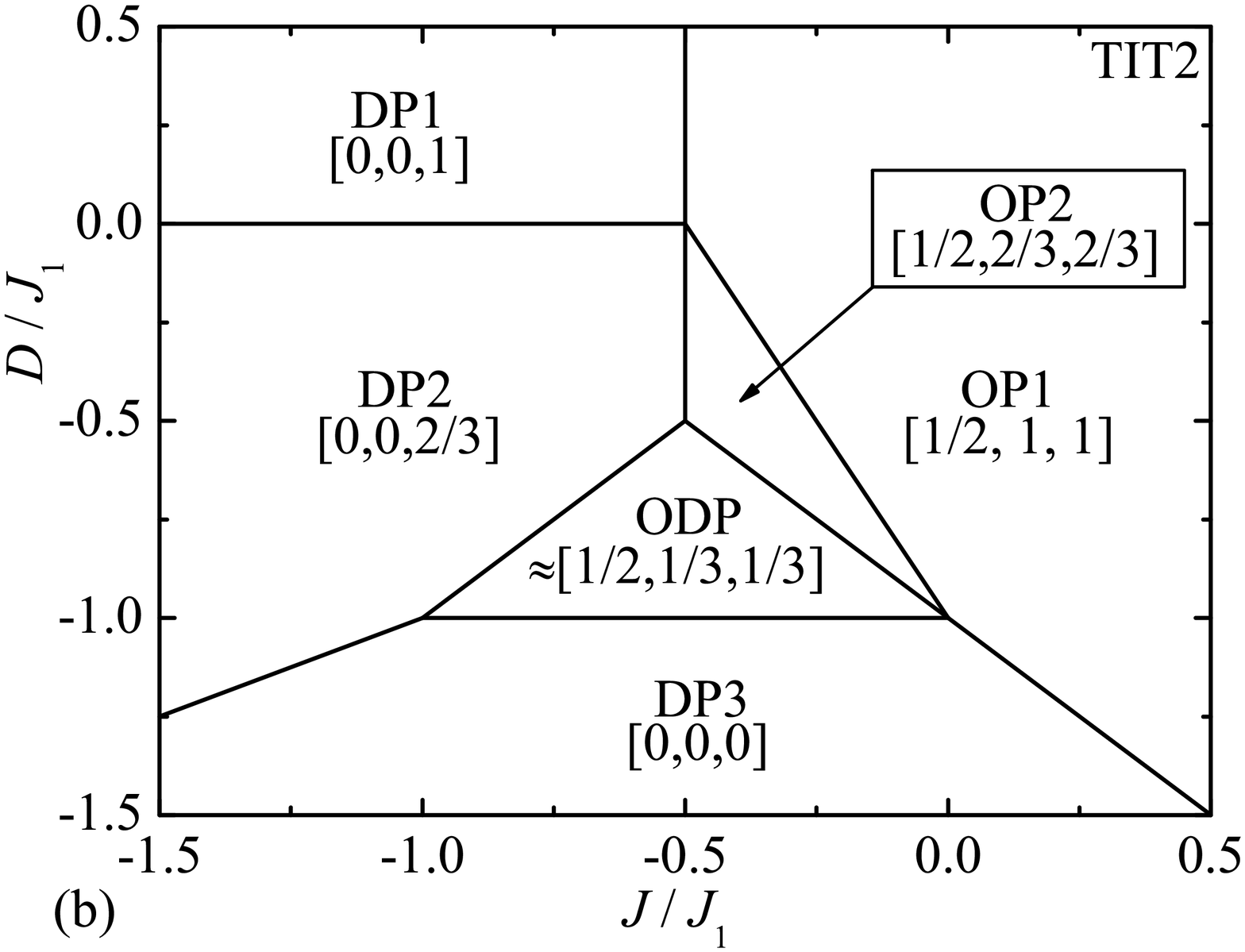}
\vspace{-0.9cm}
\caption{Ground-state phase diagrams of the mixed spin-($\frac{1}{2}$, 1) Ising model on two considered TIT lattices. Both phase diagrams are different only in a character of the ODP phase, which is critical (spontaneously ordered) on the TIT1 (TIT2) lattice. A dark-shaded (orange) area in Fig.~\ref{fig2}(a) corresponds to the parameter space, above which the spontaneous order is thermally induced by the order-from-disorder effect. The numbers in square brackets determine in a respective order the spontaneous magnetization of the nodal spins $m_{\sigma}$, the spontaneous magnetization $m_S$ and quadrupolar moment $q_S$ of the decorating spins.}
\label{fig2}
\end{figure}

To begin with, let us specify all available ground states of the mixed spin-($\frac{1}{2}$, 1) Ising model on the two considered TIT lattices. By inspection, one finds in total two spontaneously long-range ordered phases (OP1 and OP2), three disordered phases (DP1, DP2 and DP3) and one remarkable ordered-disordered phase (ODP) with a possible coexistence of the spontaneous order with a partial disorder. The ground-state phase diagram of the mixed spin-($\frac{1}{2}$, 1) Ising model on the two investigated TIT lattices, which is displayed in Fig.~\ref{fig2}, shows a region of stability of the relevant ground states. The individual ground states can be easily discerned according to zero-temperature values of the effective coupling $\beta \gamma J_{\rm eff}$, the spontaneous magnetization $m_{\sigma}$ of the nodal spins, the spontaneous magnetization $m_S$ and quadrupolar moment $q_S$ of the decorating spins, which are listed in Table \ref{tab1} for all available ground states as calculated from the respective asymptotic limits of Eqs. (\ref{TP}), (\ref{mA}), (\ref{mB}) and (\ref{qB}). 
\begin{table}
\begin{center}
\begin{tabular}{||c||c|c|c|c||}
\hline & $m_{\sigma}$ & $m_S$ & $\quad q_S \quad$ & $\beta \gamma J_{\rm eff}$  \\
\hline \hline \mbox{OP1} & $\frac{1}{2}$ & $1$ & $1$  & $\quad \infty \quad$   \\ 
\mbox{OP2} & $\frac{1}{2}$ & $\frac{2}{3}$ & $\frac{2}{3}$ & $\infty$   \\ 
\mbox{DP1} & $0$ & $0$ & $1$ & $0$  \\ 
\mbox{DP2} & $0$ & $0$ & $\frac{2}{3}$ & $-\infty$  \\ 
\mbox{DP3} & $0$ & $0$ & $0$ & $0$  \\ 
\mbox{ODP (TIT1)} & $0$ & $0$ & $\frac{1}{3}$ & $\ln 3$ \\ 
\mbox{ODP (TIT2)} & $\frac{1}{2} \sqrt[8]{\frac{125}{128}}$ & $\frac{3 \sqrt{6}-2}{16} \sqrt[8]{\frac{125}{128}}$ & $\frac{1}{3}$ & $\ln 9$  \\  
\hline  
\end{tabular}
\end{center}
\vspace{-0.4cm}
\caption{An enumeration of available ground states according to the zero-temperature values of the spontaneous magnetization $m_{\sigma}$ of the nodal spins, the spontaneous magnetization $m_S$ and quadrupolar moment $q_S$ of the decorating spins and the effective coupling $\beta \gamma J_{\rm eff}$.}
\label{tab1}
\end{table}

Let us briefly describe spin arrangements of all possible ground states. The spontaneously ordered ground state OP1 can be characterized by a perfect ferromagnetic spin arrangement of all decorating as well as nodal spins, which is realized whenever the conditions $\frac{J}{J_1}>-\frac{1}{2}$, $\frac{D}{J_1}>-1-2\frac{J}{J_1}$ and $\frac{D}{J_1}>-1-\frac{J}{J_1}$ are simultaneously satisfied. The other ordered ground state OP2 also shows a ferromagnetic spontaneous long-range order, but one out of three decorating Ising spins per decorating triangle becomes non-magnetic as it is obvious from the relevant values of the spontaneous magnetization $m_S$ and quadrupolar moment $q_S$ of the decorating spins. It is noteworthy, moreover, that the latter spontaneously ordered ground state OP2 emerges just in a rather restricted parameter region bounded by the inequalities $\frac{J}{J_1}>-\frac{1}{2}$, $\frac{D}{J_1}<-1-2\frac{J}{J_1}$ and $\frac{D}{J_1}>-1-\frac{J}{J_1}$. If the conditions $\frac{J}{J_1}<-\frac{1}{2}$ and $\frac{D}{J_1}>0$ are fulfilled, the investigated mixed-spin system captures the disordered ground state DP1 in which the nodal and decorating spins randomly occupy the spin states $\sigma_l = \pm \frac{1}{2}$ and $S_i = \pm 1$, respectively. Another disordered ground state DP2 can be detected in a parameter space delimited by $\frac{D}{J_1}<0$, $\frac{J}{J_1}<-\frac{1}{2}$, $\frac{D}{J_1}>\frac{J}{J_1}$ and $\frac{D}{J_1}>\frac{1}{2}(\frac{J}{J_1} - 1)$, where all available spin states of the nodal and decorating spins $\sigma_l = \pm \frac{1}{2}$ and $S_i = \pm 1, 0$ are occupied with the same probability. The last disordered ground state DP3 appears on assumption that $\frac{D}{J_1}<-1-\frac{J}{J_1}$, $\frac{D}{J_1}<-1$ and $\frac{D}{J_1}<\frac{1}{2}(\frac{J}{J_1} - 1)$. Under these circumstances, the complete randomness of the nodal spins occurs on behalf of the non-magnetic nature of all decorating spins, which is convincingly corroborated by the zero quadrupolar moment of the decorating spins. Finally, the most peculiar spin arrangement corresponds to the ordered-disordered ground state ODP, which emerges in a parameter space allocated by the inequalities $\frac{D}{J_1}<-1-\frac{J}{J_1}$, $\frac{D}{J_1}<\frac{J}{J_1}$ and $\frac{D}{J_1}>-1$. As evidenced by the zero-temperature limit of the effective coupling $\beta \gamma J_{\rm eff}$ listed in Table \ref{tab1}, the ground state ODP of the mixed spin-($\frac{1}{2}$, 1) Ising model on the TIT1 lattice is critical since the effective coupling of the corresponding spin-$\frac{1}{2}$ Ising model on a triangular lattice equals exactly to its critical value and consequently, the spontaneous magnetizations of the nodal and decorating spins just disappear. Contrary to this, the same ground state ODP of the mixed spin-($\frac{1}{2}$, 1) Ising model on the TIT2 lattice still exhibits a spontaneous long-range order with regard to two times greater value of the effective coupling of the corresponding spin-$\frac{1}{2}$ Ising model on a triangular lattice. However, the finite value of the effective coupling for the ODP implies a striking coexistence of the spontaneous order with a partial disorder even at zero temperature, which is in agreement with the unsaturated asymptotic values of the spontaneous magnetizations $m_{\sigma} \simeq 0.49852$ and $m_S \simeq 0.33329$ of the nodal and decorating spins, respectively (see Table \ref{tab1} for the precise asymptotic results). The exact result for the quadrupolar moment  would suggest that two decorating spins per each decorating triangle become non-magnetic, while there is almost perfect spontaneous ferromagnetic order of the nodal spins and the remaining part (one third) of the decorating spins.  
 
\begin{figure}[t]
\includegraphics[width=0.57\textwidth]{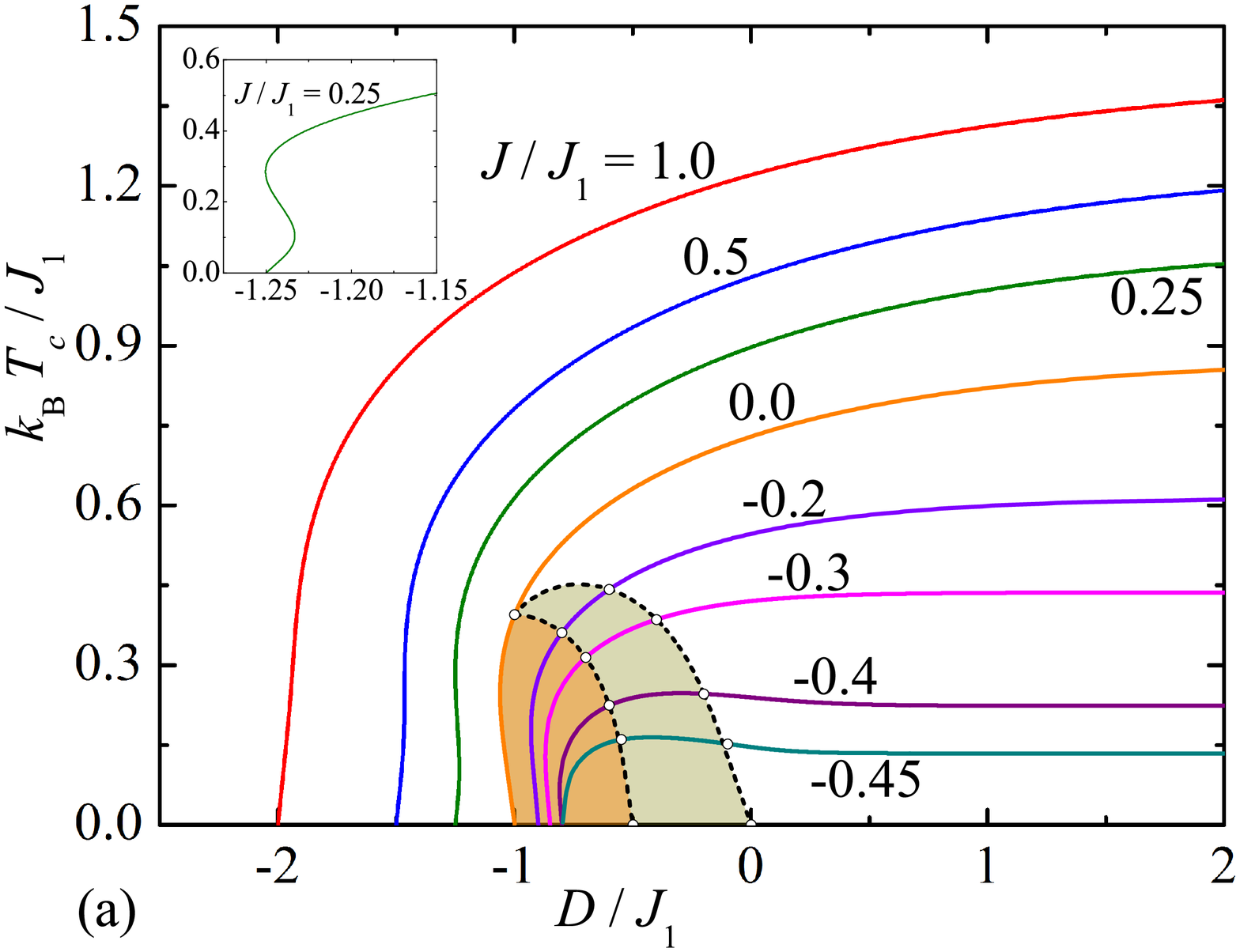}
\hspace{-1.0cm}
\includegraphics[width=0.57\textwidth]{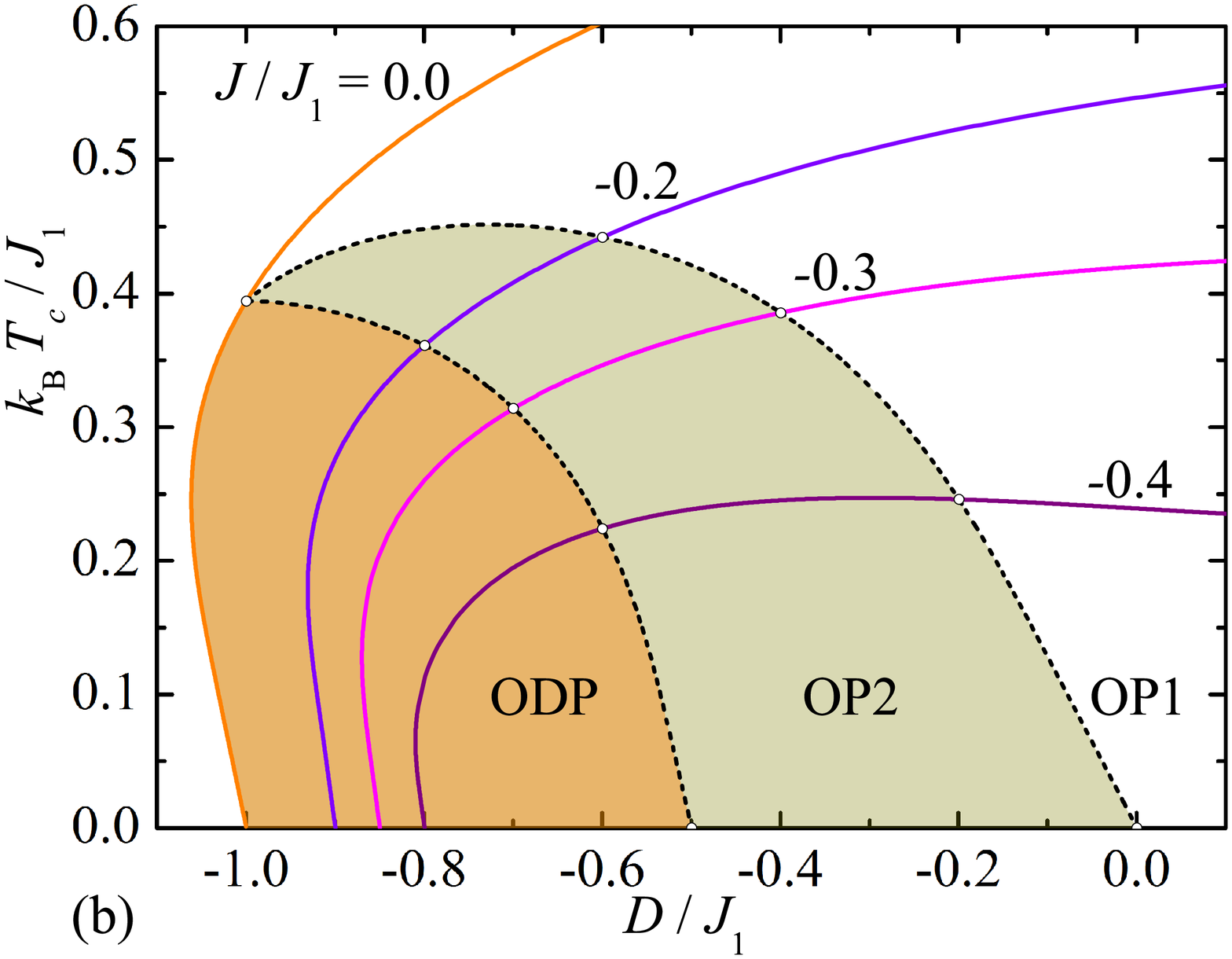}
\vspace{-0.9cm}
\caption{A critical temperature of the mixed spin-($\frac{1}{2}$, 1) Ising model on the TIT1 lattice as a function of the single-ion anisotropy for several values of the interaction ratio $\frac{J}{J_1}$. Fig. \ref{fig3}(b) shows in an enlargened scale the part of Fig. \ref{fig3}(a), which involves the OP1, OP2, and ODP phases. A part of critical lines passing through an unshaded white region corresponds to critical points of the OP1, a part of critical lines passing through a light-shaded (yellow) region corresponds to critical points of the OP2 and a part of critical lines passing through a dark-shaded (orange) region corresponds to critical points of the ODP. Inset of Fig. \ref{fig3}(a) displays a particular case with three consecutive reentrant phase transitions.}
\label{fig3}
\end{figure}

Next, we will turn our attention to a detailed analysis of the critical behaviour of the mixed spin-($\frac{1}{2}$, 1) Ising model on the two considered TIT lattices. Fig.~\ref{fig3} displays the critical temperature of the mixed spin-($\frac{1}{2}$, 1) Ising model on the TIT1 lattice as a function of the relative strength of the single-ion anisotropy $\frac{D}{J_1}$ for several values of the interaction ratio $\frac{J}{J_1}$. If the nearest-neighbour decorating spins are coupled by the ferromagnetic interaction $J>0$, the relevant critical lines always separate the usual ferromagnetically ordered phase OP1 from the disordered paramagnetic phase. It should be nevertheless noticed that the monotonous dependences of the critical temperature are observed just for $\frac{J}{J_1} \geq \frac{1}{2}$, while more intriguing dependences of the critical temperature indicating an existence of reentrant phase transitions with three consecutive critical points emerge for $\frac{1}{2} > \frac{J}{J_1} > 0$ (see the inset in Fig.~\ref{fig3}(a)). On the other hand, the critical behaviour becomes much more intricate for the antiferromagnetic interaction between the nearest-neighbour decorating spins $J<0$, which is for better clarity depicted in Fig.~\ref{fig3}(b) in an enlargened scale. Under this condition, a part of critical lines passing through the unshaded (white) region corresponds to the critical points of OP1, a part of critical lines passing through the light-shaded (yellow) region corresponds to the critical points of OP2 and a part of critical lines passing through the dark-shaded (orange) region is pertinent to the critical points of ODP. Thus, it surprisingly turns out that the temperature may induce the spontaneous long-range order just above the critical ground state ODP through the order-from-disorder effect, but only in a rather restricted parameter space given by $-1-\frac{J}{J_1}>\frac{D}{J_1}>-1-\frac{1}{2} \frac{J}{J_1}$ that is highlighted in the ground-state phase diagram (Fig.~\ref{fig2}(a)) by the dark-shaded (orange) region. It is also quite clear from Fig.~\ref{fig3}(b) that the antiferromagnetic coupling between the nearest-neighbour decorating spins $J < 0$ leads in combination with a sufficiently strong but not too strong easy-plane single-ion anisotropy $\frac{D}{J_1} \lessapprox -1-\frac{1}{2} \frac{J}{J_1}$ to an emergence of another kind of reentrant phase transition with two successive critical points.  
   
\begin{figure}[t]
\includegraphics[width=0.57\textwidth]{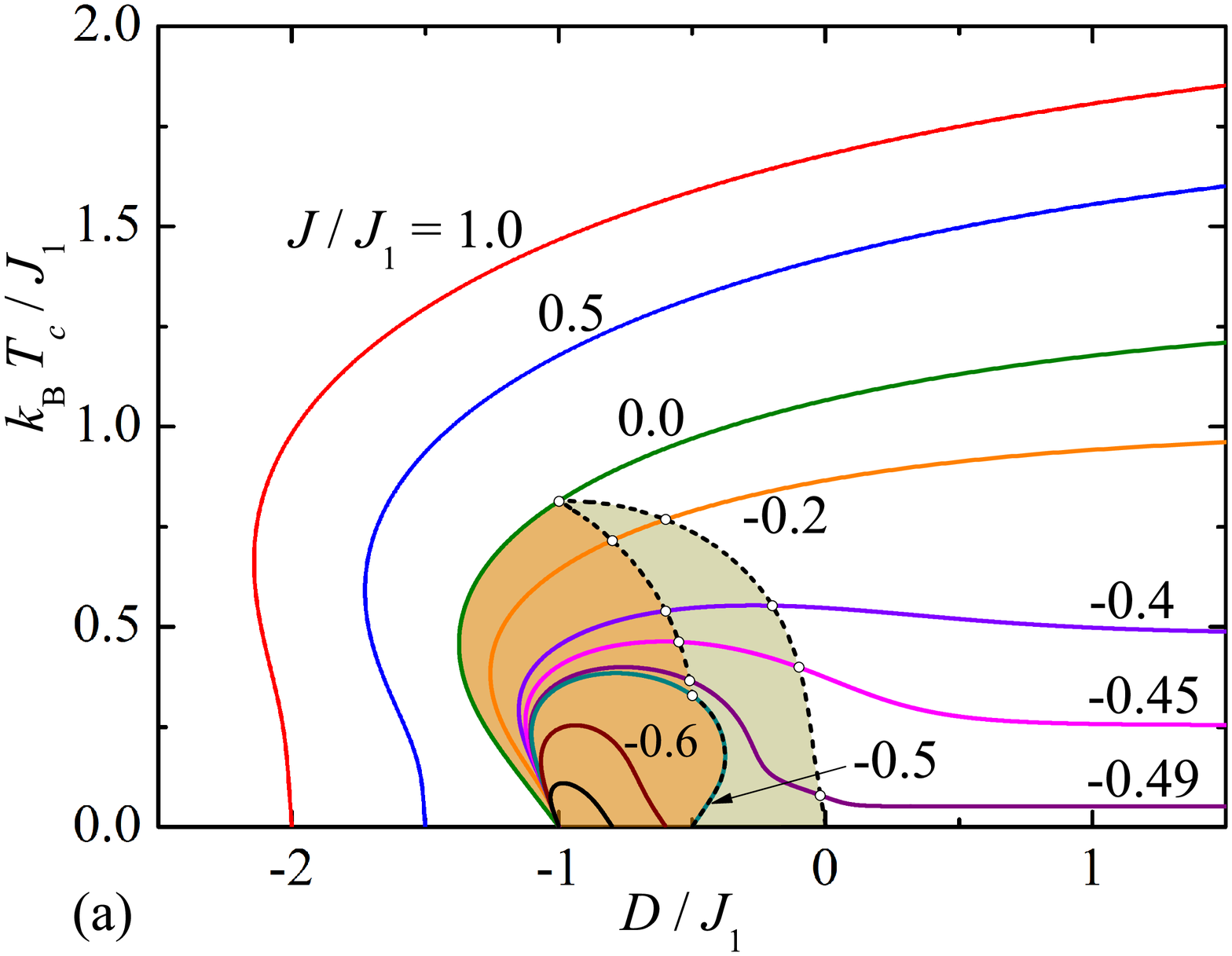}
\hspace{-1.0cm}
\includegraphics[width=0.57\textwidth]{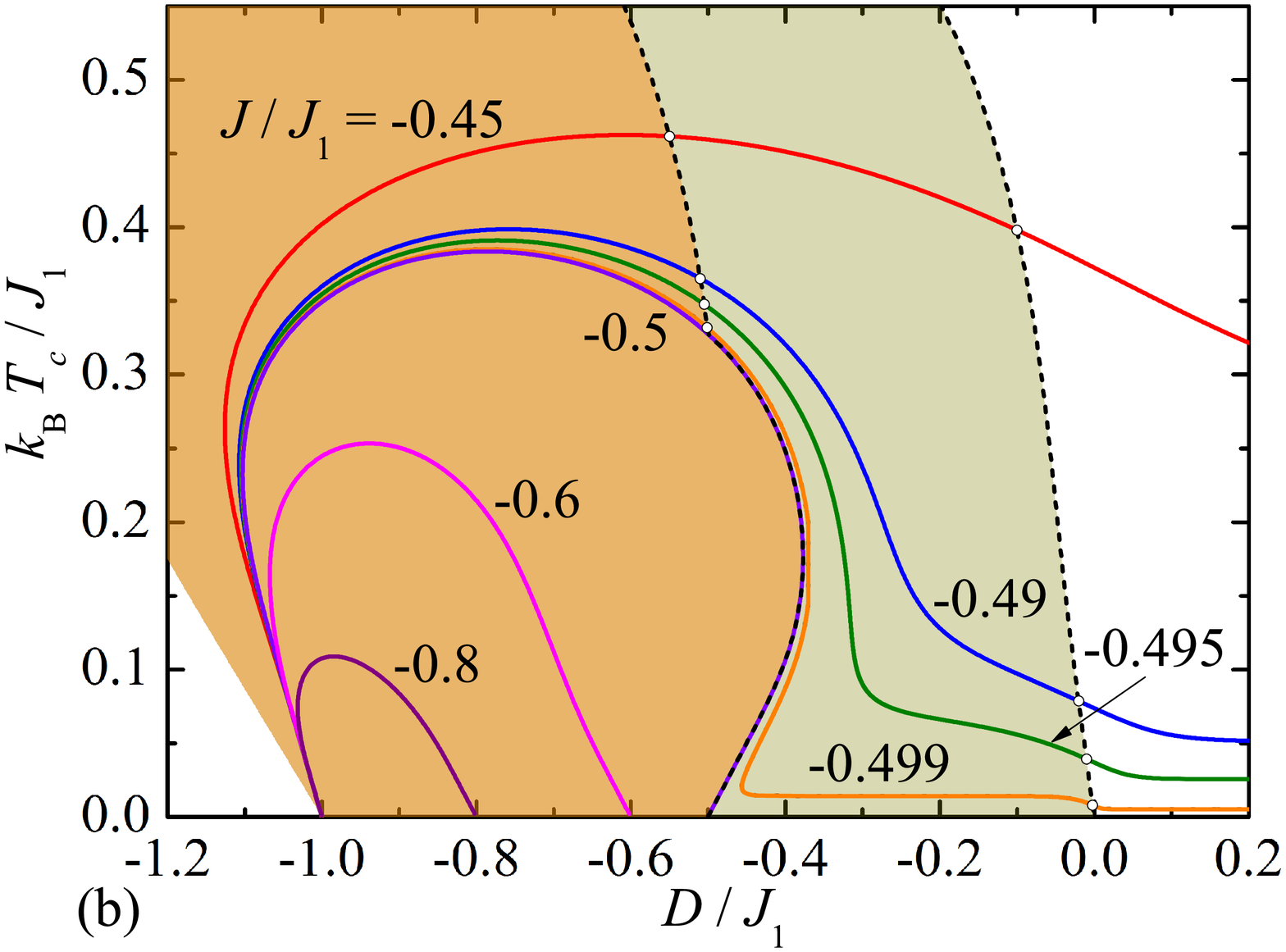}
\vspace{-0.9cm}
\caption{A critical temperature of the mixed spin-($\frac{1}{2}$, 1) Ising model on the TIT2 lattice as a function of the single-ion anisotropy for several values of the interaction ratio $\frac{J}{J_1}$. Fig. \ref{fig4}(b) shows in an enlargened scale the part of Fig. \ref{fig4}(a), which involves the OP1, OP2, and ODP phases. A part of critical lines passing through an unshaded white region corresponds to critical points of the OP1, a part of critical lines passing through a light-shaded (yellow) region corresponds to critical points of the OP2 and a part of critical lines passing through a dark-shaded (orange) region corresponds to critical points of the ODP.}
\label{fig4}
\end{figure}

For comparison, the critical temperature of the mixed spin-($\frac{1}{2}$, 1) Ising model on the TIT2 lattice is plotted in Fig.~\ref{fig4} against a relative strength of the single-ion anisotropy $\frac{D}{J_1}$ for different values of the interaction ratio $\frac{J}{J_1}$. Altogether, the most crucial difference in the critical behaviour of the two considered TIT lattices can be viewed in a greater resistance of the TIT2 lattice with respect to the spin frustration. As a matter of fact, the mixed spin-($\frac{1}{2}$, 1) Ising model on the TIT2 lattice becomes firmly disordered merely for the stronger antiferromagnetic interactions between the decorating spins $\frac{J}{J_1} < -1$ than the analogous model on the TIT1 lattice being persistently disordered for any $\frac{J}{J_1} < -\frac{1}{2}$. A greater stability of the mixed spin-($\frac{1}{2}$, 1) Ising model on the TIT2 lattice against the spin frustration can be attributed to the nature of ODP, which is spontaneously long-range ordered for the model defined on the TIT2 lattice but is critical for the analogous model on the TIT1 lattice. Apart from this fact, the critical lines of the mixed spin-($\frac{1}{2}$, 1) Ising model on the TIT2 lattice exhibit more pronounced reentrant phase transitions with two consecutive critical points, which occur on assumption that the single-ion anisotropy is selected sufficiently close but slightly below: (i) the ground-state phase boundary between OP1 and DP3 ($\frac{D}{J_1} \lessapprox -1-\frac{J}{J_1}$ for $\frac{J}{J_1} > 0$); (ii) the ground-state phase boundary between ODP and DP3 ($\frac{D}{J_1} \lessapprox -1$ for $0 > \frac{J}{J_1} > -1$). If the interaction parameters are selected close enough but slightly above the ground-state boundary between OP2 and ODP, i.e. $\frac{D}{J_1} \gtrapprox -0.5$ and $\frac{J}{J_1} \gtrapprox -0.5$, reentrant phase transitions with three successive critical points can be detected. This reentrance reflects three consecutive temperature-induced phase transitions from the spontaneously ordered phase OP2 to the disordered paramagnetic phase, the disordered paramagnetic phase to the partially ordered and partially disordered phase ODP and vice versa.

\begin{figure}[t]
\includegraphics[width=0.57\textwidth]{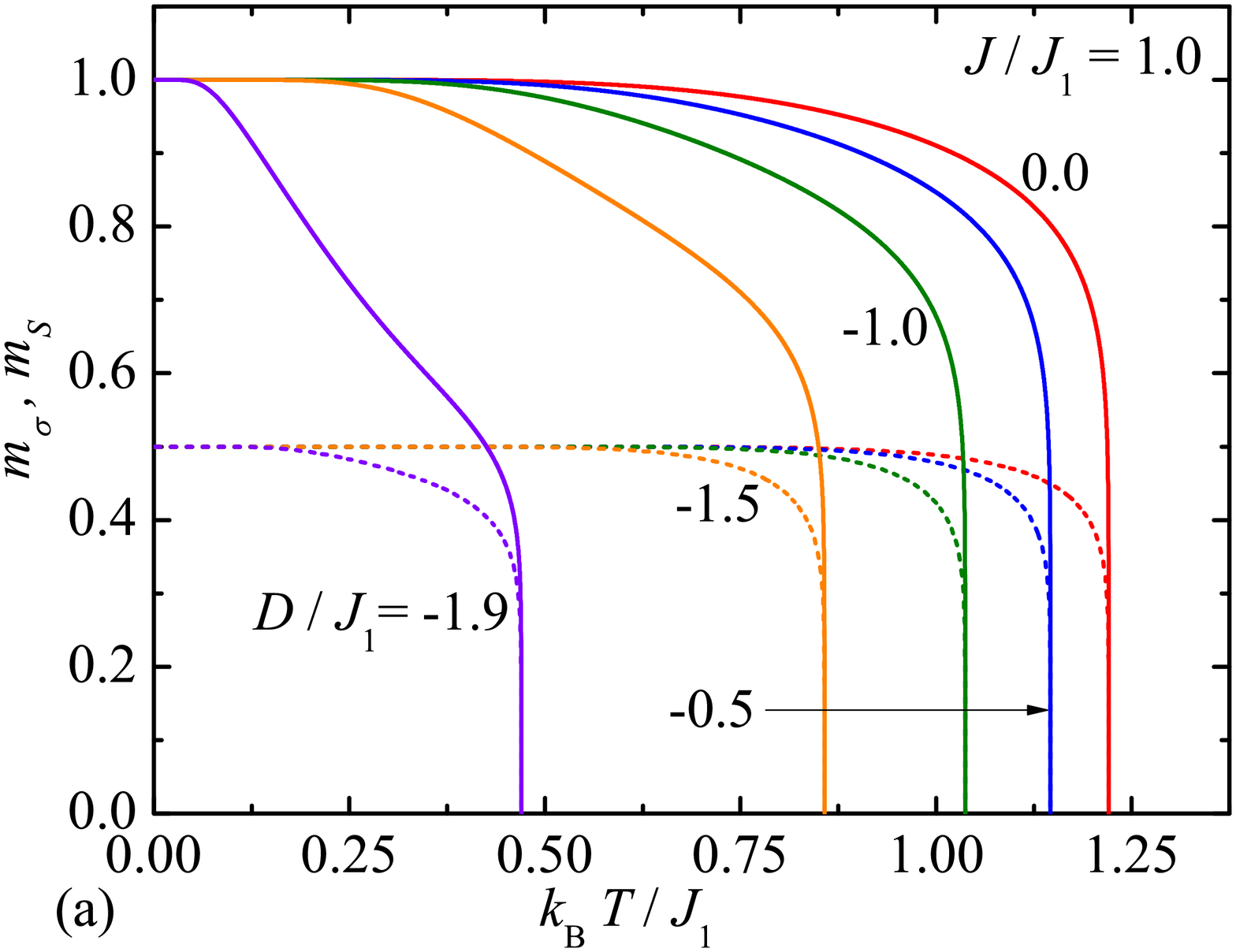}
\hspace{-1.0cm}
\includegraphics[width=0.57\textwidth]{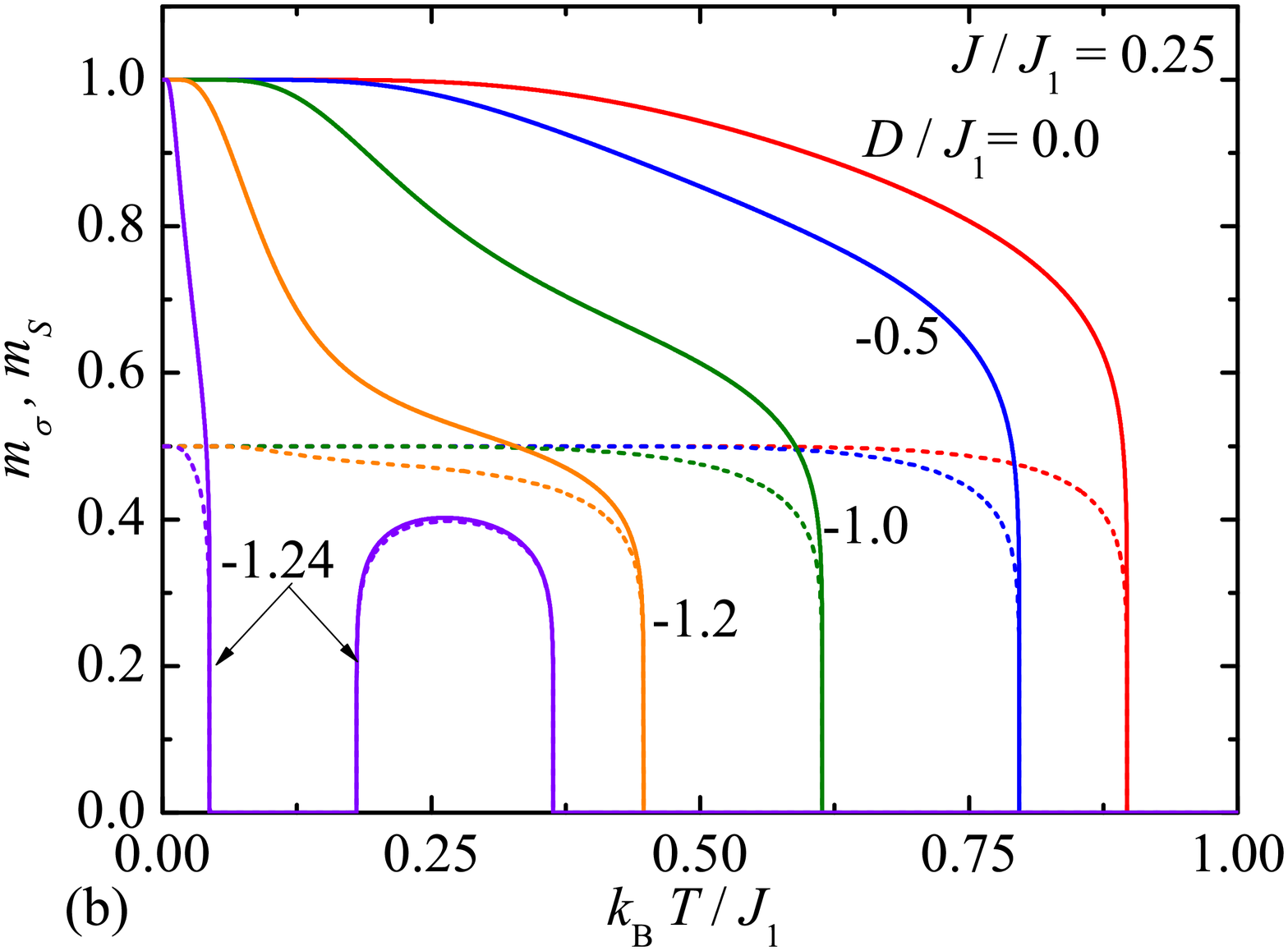}
\vspace{-0.9cm}
\caption{Thermal variations of the spontaneous magnetization of the mixed spin-($\frac{1}{2}$, 1) Ising model on the TIT1 lattice for several values of the single-ion anisotropy and two different values of the interaction ratio: (a) $\frac{J}{J_1} = 1$; (b) $\frac{J}{J_1} = \frac{1}{4}$. Solid (broken) lines show the spontaneous magnetization of the decorating (nodal) spins, respectively.}
\label{fig5}
\end{figure}

In the following, we will confirm the unexpectedly diverse critical behaviour reported previously through an investigation of typical thermal dependences of the spontaneous magnetization. For this purpose, temperature variations of the spontaneous magnetization of the nodal and decorating spins are displayed in Fig.~\ref{fig5} for the mixed spin-($\frac{1}{2}$, 1) Ising model on the TIT1 lattice with the ferromagnetic coupling between the nearest-neighbour decorating spins. Fig.~\ref{fig5}(a) illustrates temperature dependences of the spontaneous magnetizations typical for sufficiently strong ferromagnetic interactions $\frac{J}{J_1} \geq \frac{1}{2}$. Under this condition, the spontaneous magnetization $m_S$ of the decorating spins becomes more subtle with respect to thermal fluctuations upon strengthening of the single-ion anisotropy, which results in a downward curvature observable at moderate temperatures caused by a thermal population of the non-magnetic state $S_i=0$. On the other hand, the other particular case shown in Fig.~\ref{fig5}(b) illustrates typical temperature dependences of the spontaneous magnetizations for weaker ferromagnetic interactions $\frac{1}{2} > \frac{J}{J_1} > 0$. It could be concluded that one observes the same general trends as before with exception of a peculiar temperature dependence of the spontaneous magnetization, which evidences reentrant phase transitions with three successive critical points for $\frac{D}{J_1} \gtrapprox -1 - \frac{J}{J_1}$. 

\begin{figure}[t]
\includegraphics[width=0.57\textwidth]{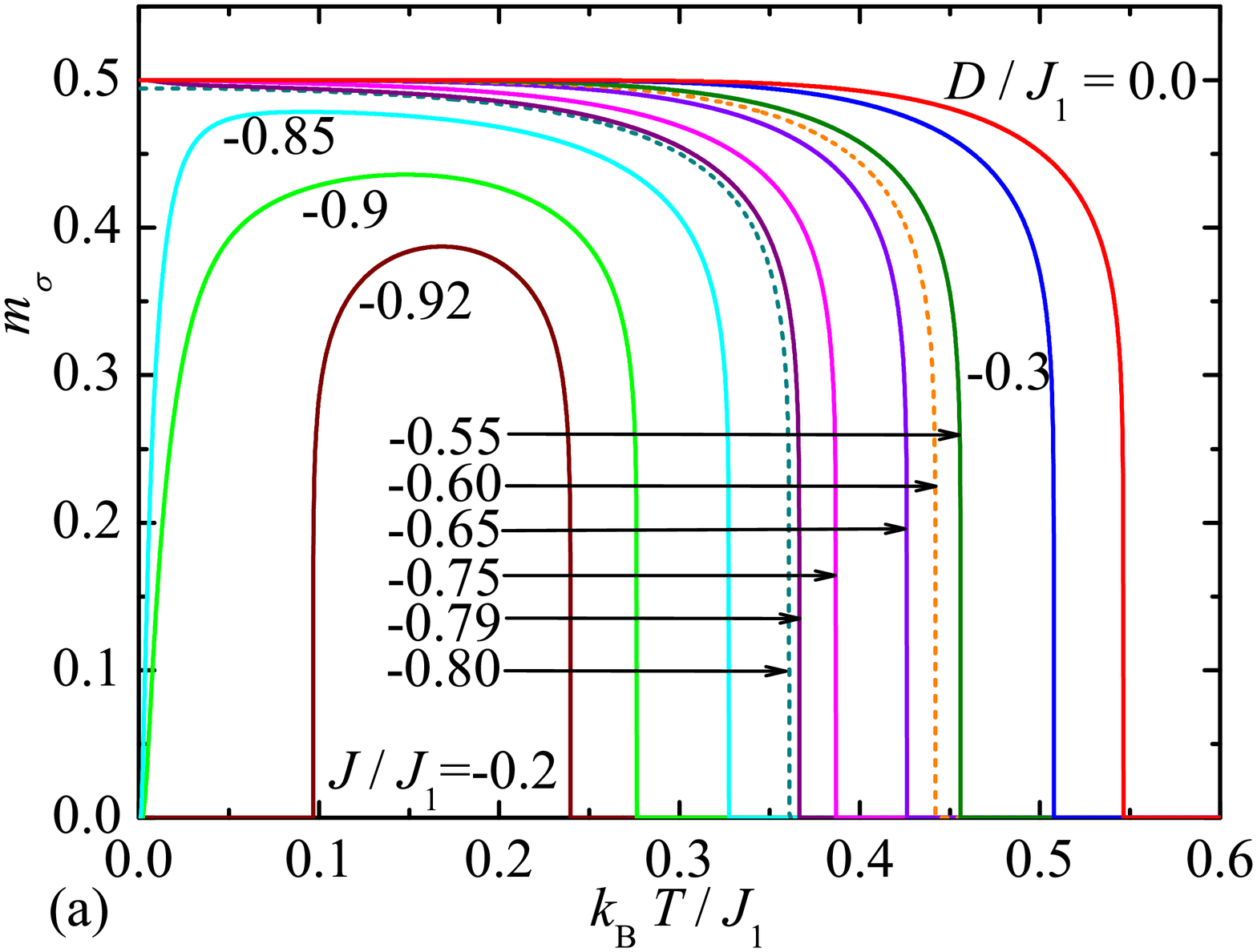}
\hspace{-1.0cm}
\includegraphics[width=0.57\textwidth]{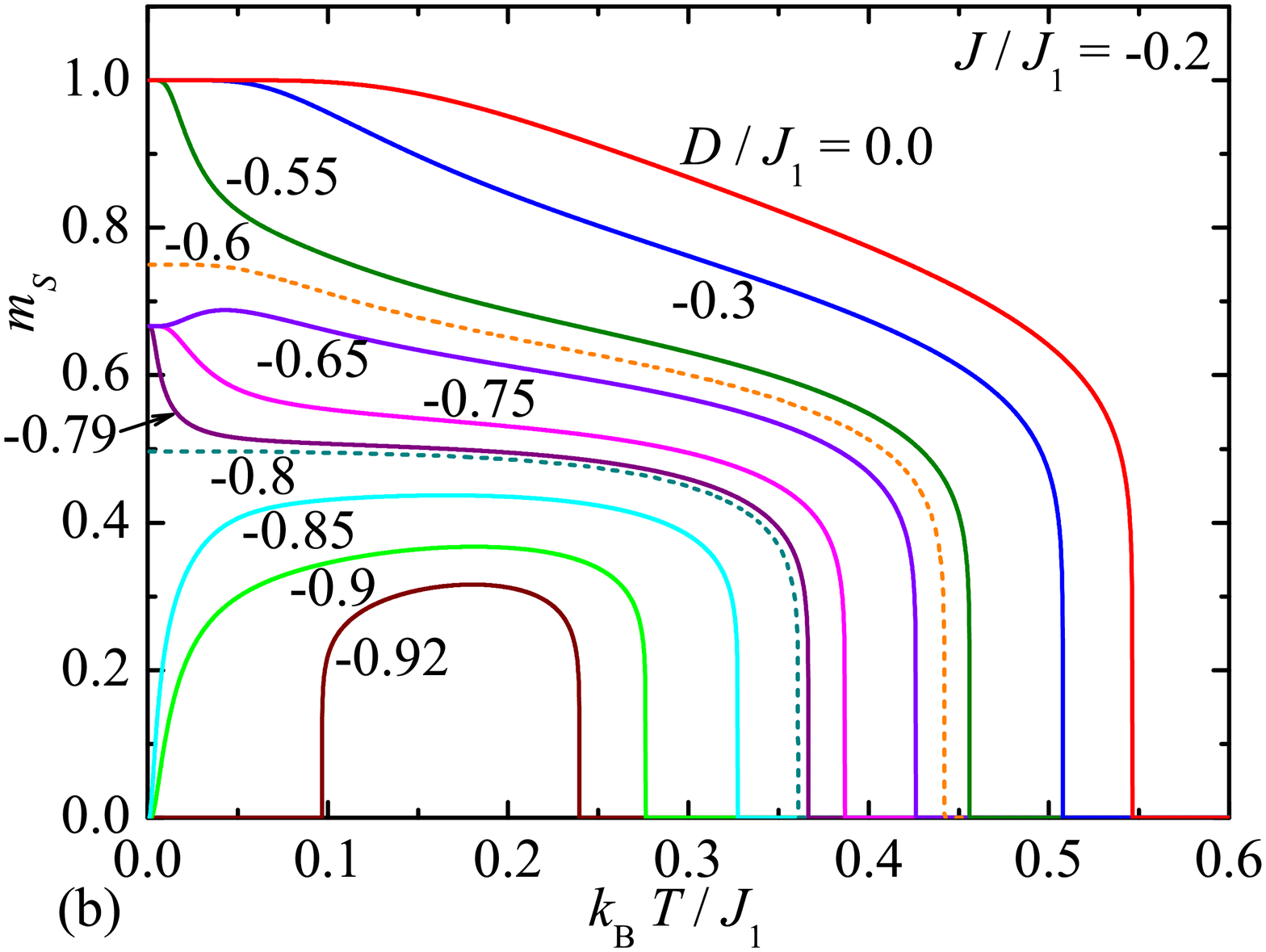}
\vspace{-0.9cm}
\caption{Thermal variations of the spontaneous magnetization of the mixed spin-($\frac{1}{2}$, 1) Ising model on the TIT1 lattice for one fixed value of the interaction ratio $\frac{J}{J_1} = -\frac{1}{5}$ and several values of the single-ion anisotropy. Fig.~\ref{fig6}(a) (Fig.~\ref{fig6}(b)) shows the spontaneous magnetization of the nodal (decorating) spins. Broken lines corresponds to special points of phase coexistence.}
\label{fig6}
\end{figure}

Furthermore, let us bring insight into typical thermal dependences of the spontaneous magnetization of the mixed spin-($\frac{1}{2}$, 1) Ising model on the TIT1 lattice with the antiferromagnetic coupling between the nearest-neighbour decorating spins. For illustration, the spontaneous magnetization of the nodal and decorating spins is plotted in Fig.~\ref{fig6}(a) and Fig.~\ref{fig6}(b) against the temperature for one special value of the interaction ratio $\frac{J}{J_1} = -\frac{1}{5}$. In agreement with the finite-temperature phase diagram shown in Fig.~\ref{fig3}, the spontaneous magnetization of the nodal and decorating spins achieve at low enough temperatures their saturated values only if $\frac{D}{J_1}>-1-2\frac{J}{J_1}$. This result is consistent with a stability region of the fully saturated ferromagnetic ground state OP1. If the easy-plane single-ion anisotropy of moderate strength $-1-2\frac{J}{J_1}>\frac{D}{J_1}>-1-\frac{J}{J_1}$ is selected, however, the spontaneous magnetization of the decorating spins then converges at sufficiently low temperatures to the asymptotic value $\frac{2}{3}$ that is reminiscent of the OP2 with regard to the full saturation of the spontaneous magnetization of the nodal spins. It is worthwhile to remark, moreover, that the spontaneous magnetizations rising from zero serve in evidence of a peculiar spontaneous order, which can be thermally induced above the critical ground state ODP if the single-ion anisotropy is chosen from the range $-1-\frac{J}{J_1}>\frac{D}{J_1}>-1-\frac{1}{2}\frac{J}{J_1}$. Finally, it should be emphasized that thermal variations of the spontaneous magnetization verify an appearance of reentrant phase transitions with two different non-zero critical temperatures provided that $\frac{D}{J_1} \lessapprox -1-\frac{1}{2}\frac{J}{J_1}$.    
 
\begin{figure}[t]
\includegraphics[width=0.57\textwidth]{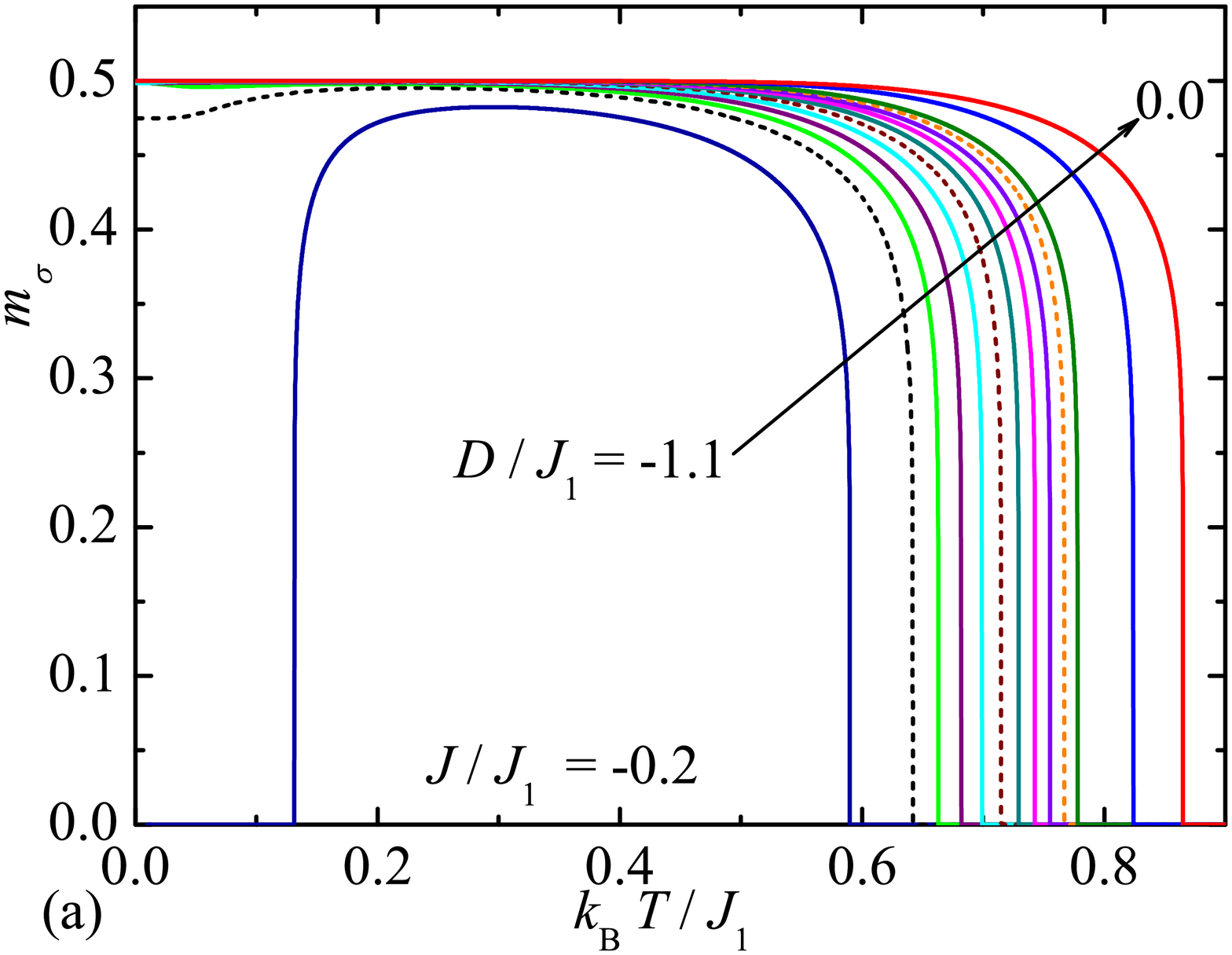}
\hspace{-1.0cm}
\includegraphics[width=0.57\textwidth]{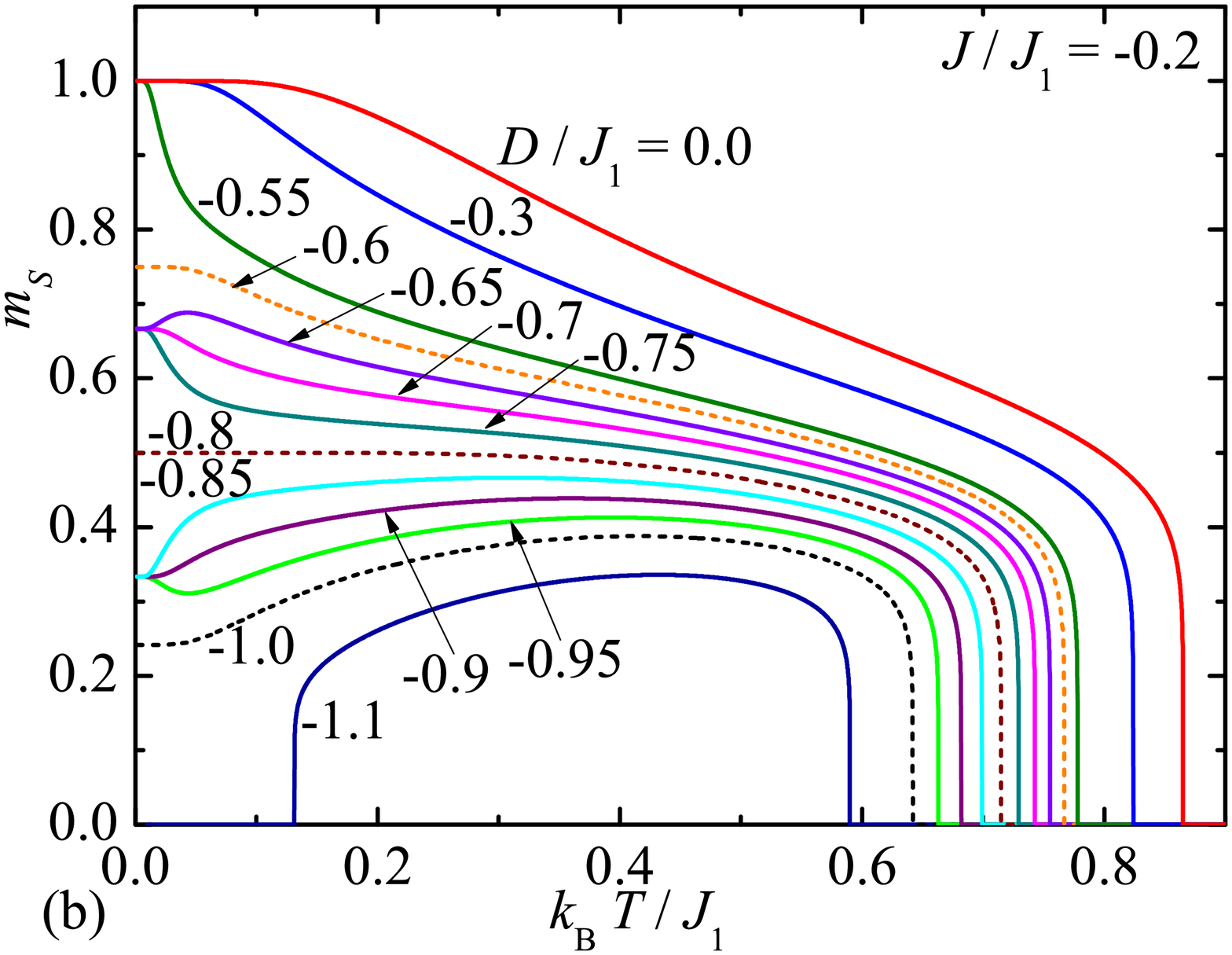}
\vspace{-0.9cm}
\caption{Thermal variations of the spontaneous magnetization of the mixed spin-($\frac{1}{2}$, 1) Ising model on the TIT2 lattice for one fixed value of the interaction ratio $\frac{J}{J_1} = -\frac{1}{5}$ and several values of the single-ion anisotropy. Fig.~\ref{fig7}(a) (Fig.~\ref{fig7}(b)) shows the spontaneous magnetization of the nodal (decorating) spins. The single-ion anisotropy changes in Fig.~\ref{fig7}(a) in a ascending order along the direction of arrow using the same values as in Fig.~\ref{fig7}(b).}
\label{fig7}
\end{figure}

\begin{figure}[t]
\includegraphics[width=0.57\textwidth]{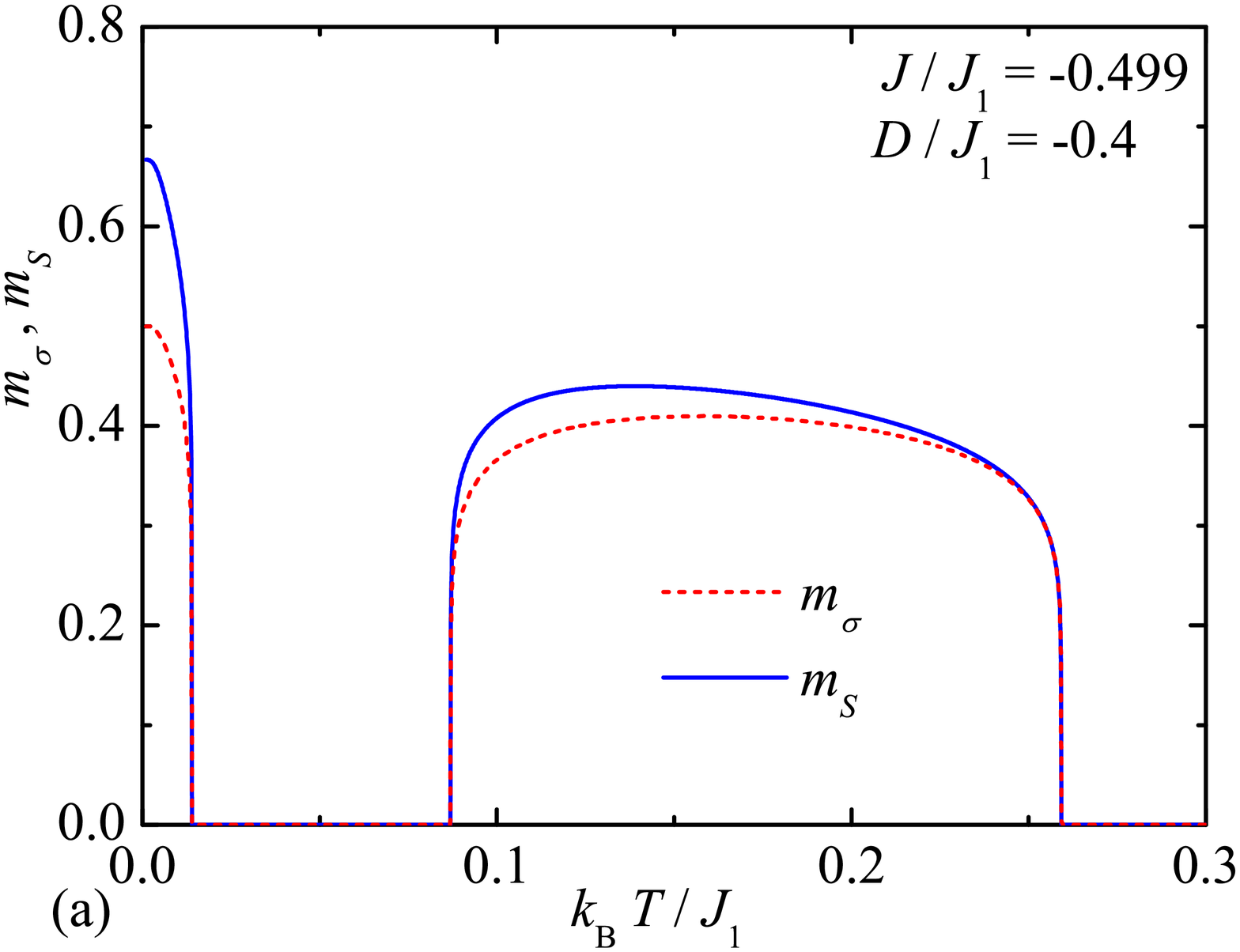}
\hspace{-1.0cm}
\includegraphics[width=0.57\textwidth]{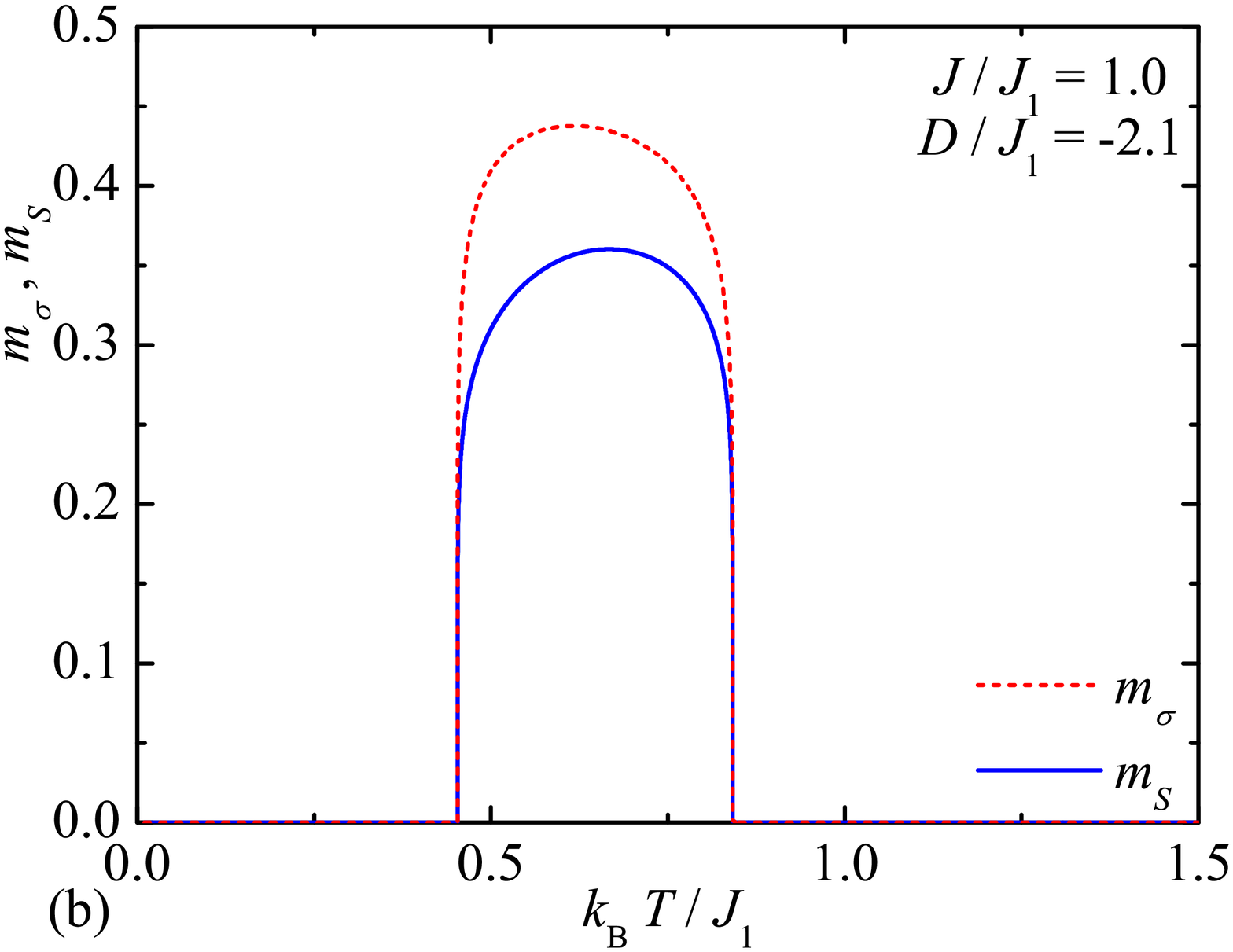}
\vspace{-0.9cm}
\caption{Temperature dependences of the spontaneous magnetizations of the mixed spin-($\frac{1}{2}$, 1) Ising model on the TIT2 lattice for two different sets of the interaction parameters: (a) $\frac{J}{J_1} = -0.499$, $\frac{D}{J_1} = -0.4$; (b) $\frac{J}{J_1} = 1.0$, $\frac{D}{J_1} = -2.1$.}
\label{fig8}
\end{figure}

Last but not least, let us examine a few outstanding temperature dependences of the spontaneous magnetization of the mixed spin-($\frac{1}{2}$, 1) Ising model on the TIT2 lattice, which cannot be in principle found in the analogous Ising model defined on the TIT1 lattice. First, it is of particular interest to verify the ordered nature of ODP at zero temperature, which is relevant just for the mixed-spin Ising model on the TIT2 lattice on assumption that the nearest-neighbour interaction between the decorating spins is antiferromagnetic. It is quite evident from Fig.~\ref{fig7} that the spontaneous magnetizations of the nodal and decorating spins indeed tend towards non-zero asymptotic values in the parameter space ascribed to the ODP ($-1-\frac{J}{J_1}>\frac{D}{J_1}>-1$) as temperature goes to zero. In addition, the respective asymptotic values of both spontaneous magnetizations are in accordance with the exact results reported in Table~\ref{tab1} for the ODP ($m_{\sigma} \simeq 0.49852$ and $m_S \simeq 0.33329$). Second, Fig.~\ref{fig8}(a) proves an existence of three successive reentrant phase transitions of the mixed-spin Ising model on the TIT2 lattice in a vicinity of the ground-state phase boundary between OP2 and ODP. The proposed mechanism for the triple reentrance including continuous phase transitions between the spontaneously ordered phase OP2 and the disordered paramagnetic phase, the disordered paramagnetic phase and the ordered-disordered phase ODP and vice versa, are fully consistent with the appropriate values of the spontaneous magnetizations of the nodal and decorating spins. Third, the double reentrance with two successive phase transitions is illustrated in Fig.~\ref{fig8}(b) for the mixed spin-($\frac{1}{2}$, 1) Ising model on the TIT2 lattice with a relatively strong ferromagnetic coupling between the nearest-neighbour decorating spins. It should be remembered that this kind of double reentrance can be found in the analogous mixed-spin Ising model on the TIT1 lattice only if the nearest-neighbour decorating spins are coupled by a rather weak ferromagnetic interaction $\frac{1}{2}>\frac{J}{J_1}>0$.    

\section{Conclusion}
\label{conclusion}

The present article deals with the mixed spin-($\frac{1}{2}$, 1) Ising model on two geometrically frustrated TIT lattices, which has been exactly solved through the generalized star-triangle transformation establishing a rigorous mapping equivalence with the corresponding spin-$\frac{1}{2}$ Ising model on a triangular lattice. Within the framework of this precise mapping correspondence, we have derived exact results for the partition function, the spontaneous magnetization of the nodal spins, the spontaneous magnetization and quadrupolar moment of the decorating spins along with the critical condition allocating critical points. It has been evidenced that the variety of ordered, disordered and ordered-disordered ground states emerge owing to a mutual interplay between the spin frustration and single-ion anisotropy, which is simultaneously responsible for a diverse critical behaviour including the order-from-disorder effect and reentrant phase transitions with either two or three successive critical points. 

Although the exact treatment of the mixed spin-($\frac{1}{2}$, 1) Ising model on two considered TIT lattices differs just in a relative strength of the effective interaction of the corresponding spin-$\frac{1}{2}$ Ising model on a triangular lattice, it turns out that this rather small technical difference might cause a significant change in a character of some ground states and the overall critical behaviour. The most outstanding finding of the present work certainly represents a possible coexistence of the spontaneous order with a partial disorder within the ODP. While a striking coexistence of the spontaneous order with disorder ODP is just thermally induced through the order-from-disorder effect in the mixed-spin Ising model on the TIT1 lattice, the analogous model defined on the TIT2 lattice exhibits this remarkable ordered-disordered state even at zero temperature. Besides, it has been also demonstrated that the latter model on the TIT2 lattice generally exhibits more pronounced reentrance extending over broader parameter space than the former model on the TIT1 lattice.


\begin{thebibliography}{100}
\bibitem{lieb86} R. Liebmann, Statistical Mechanics of Periodic Frustrated Ising Systems, Lecture Notes in Physics, Vol. 251, Springer, Berlin, 1986.  
\bibitem{diep04} H.T. Diep, Frustrated Spin Systems, World Scientific, Singapore, 2004.
\bibitem{hout50} R.M.F. Houtappel, Physica 16 (1950) 425.
\bibitem{wann50} G.H. Wannier, Phys. Rev. 79 (1950) 357; Erratum: Phys. Rev. B 7 (1972) 5017.
\bibitem{temp50} H.N.V. Temperley, Proc. R. Soc. Lond. A 202 (1950) 202.
\bibitem{husi50} K. Husimi, I. Syozi, Prog. Theor. Phys. 5 (1950) 177.
\bibitem{syoz50} I. Syozi, Prog. Theor. Phys. 5 (1950) 341.
\bibitem{newe50} G.F. Newell, Phys. Rev. 79 (1950) 876. 
\bibitem{pott52} R.B. Potts, Phys. Rev. 88 (1952) 352.
\bibitem{domb60} C. Domb, Adv. Phys. 9 (1960) 149. 
\bibitem{step70} J. Stephenson, J. Math. Phys. 11 (1970) 413.
\bibitem{cunn74} G.W. Cunningham, P.H.E. Meijer, J. Math. Phys. 15 (1974) 55.
\bibitem{baxt89} R.J. Baxter, T.C. Choy, Proc. R. Soc. Lond.  A 423 (1989) 279.
\bibitem{vaks66} V.G. Vaks, A.I. Larkin, N. Ovchinnikov, Sov. Phys. JETP 22 (1966) 820.
\bibitem{mori86} T. Morita, J. Phys. A: Math. Gen. 19 (1986) 1701.
\bibitem{chik87} T. Chikyu and M. Suzuki, Prog. Theor. Phys. 78 (1987) 1242.
\bibitem{kano53} K. Kano, S. Naya, Prog. Theor. Phys. 10 (1953) 158.
\bibitem{kano66} K. Kano, Prog. Theor. Phys. 35 (1966) 1.
\bibitem{vill77} J. Villian, J. Phys. C: Solid State Phys. 10 (1977) 1717.
\bibitem{wolf82} W.F. Wolff, J. Zittartz, Z. Phys. B 49 (1982) 139.
\bibitem{seze81} R. Bidaux, L. de Seze, J. Physique 42 (1981) 371.
\bibitem{wozi82} W.F. Wolff, J. Zittartz, Z. Phys. B 47 (1982) 341.
\bibitem{wald84} M.H. Waldor, W.F. Wolff, J. Zittartz, Phys. Lett. A 106 (1984) 261.
\bibitem{wald85} M.H. Waldor, W.F. Wolff, J. Zittartz, Z. Phys. B 59 (1985) 43.
\bibitem{zhen05} J. Zheng, G. Sun, Phys. Rev. B 71 (2005) 052408. 
\bibitem{loh08}  Y.L. Loh, D.X. Yao, E.W. Carlson, Phys. Rev. B 77 (2008) 134402.
\bibitem{andr79} G. Andr\'e, R. Bidaux, J.P. Carton, R. Conte, L. de Seze, J. Physique 40 (1979) 479.
\bibitem{diep91} H.T. Diep, M. Debauche, H. Giacomini, Phys. Rev. B 43 (1991) 8759.
\bibitem{diep92} H.T. Diep, M. Debauche, H. Giacomini, J. Magn. Magn. Mater. 104--107 (1992) 184.
\bibitem{azar87} P. Azaria, H.T. Diep, H. Giacomini, Phys. Rev. Lett. 59 (1987) 1629.
\bibitem{deba91} M. Debauche, H.T. Diep, P. Azaria, H. Giacomini, Phys. Rev. B 44 (1991) 2369.
\bibitem{vill80} J. Villian, R. Bidaux, J.P. Carton, R. Conte, J. Physique 41 (1980) 1263.
\bibitem{bida81} R. Bidaux, J.P. Carton, R. Conte, J. Villian, in Disordered Systems and Localization, 
edited by: C. Castellani, C. Di Castro, L. Peliti, Lecture Notes in Physics, Vol. 149, 1981, p. 161.
\bibitem{zitt82} W.F. Wolff, J. Zittartz, Z. Phys. B 49 (1982) 229.
\bibitem{gonc87} L.L. Gon\c{c}alves, Phys. Scripta 35 (1987) 105.
\bibitem{lipo95} A. Lipowski, T. Horiguchi, J. Phys. A: Math. Gen. 28 (1995) L261.
\bibitem{stre06} J. Stre\v{c}ka, Phys. stat. sol. (b) 243 (2006) 708.
\bibitem{cano06} J. Stre\v{c}ka, L. \v{C}anov\'a, J. Dely, Phys. stat. sol. (b) 243 (2006) 1946.
\bibitem{jasc05} M. Ja\v{s}\v{c}ur, J. Stre\v{c}ka, Condens. Matter. Phys. 8 (2005) 869.
\bibitem{stre12} J. Stre\v{cka}, C. Ekiz, Physica A 391 (2012) 4763.
\bibitem{fish59} M.E. Fisher, Phys. Rev. 113 (1959) 969. 
\bibitem{roja09} O. Rojas, J.S. Valverde and S.M. De Souza, Physica A 388 (2009) 1419.
\bibitem{stre10} J. Stre\v{c}ka, On the Theory of Generalized Algebraic Transformations, LAP LAMBERT Academic Publishing, Saarbrucken, 2010.
\bibitem{stre10v} J. Stre\v{c}ka, Phys. Lett. A 374 (2010) 3718.
\bibitem{barr82} J.H. Barry, C.H. M\'unera, T. Tanaka, Physica A 113 (1982) 367.
\bibitem{barr88} J.H. Barry, M. Khatun, T. Tanaka, Phys. Rev. B 37 (1988) 5193.
\bibitem{khat90} M. Khatun, J.H. Barry, T. Tanaka, Phys. Rev. B 42 (1990) 4398.
\bibitem{barr91} J.H. Barry, T. Tanaka, M. Khatun, C.H. M\'unera, Phys. Rev. B 44 (1991) 2595.
\bibitem{barr95} J.H. Barry, M. Khatun, Phys. Rev. B 51 (1995) 5840.
\bibitem{call63} H.B. Callen, Phys. Lett. 4 (1963) 161.
\bibitem{suzu65} M. Suzuki, Phys. Lett. 19 (1965) 267.
\bibitem{balc02} T. Balcerzak, J. Magn. Magn. Mater. 246 (2002) 213. 





\end{thebibliography}
\end{document}